\documentclass[acmlarge]{acmart}
\usepackage{booktabs} 
\usepackage{multirow}
\usepackage{array}
\usepackage{graphicx}
\usepackage[normalem]{ulem}
\useunder{\uline}{\ul}{}
\usepackage[caption=false]{subfig}
\usepackage{cleveref}
\usepackage{rotating}
\usepackage{siunitx}
\usepackage{makecell}
\usepackage{tablefootnote}
\usepackage{makecell}

\AtBeginDocument{%
  }

\setcopyright{none}
\makeatletter
\renewcommand\@formatdoi[1]{\ignorespaces}
\makeatother
\renewcommand{\footnotetextcopyrightpermission}[1]{}
\begin{document}

\title{Towards Better Understanding of User Satisfaction in Open-Domain Conversational Search}

\author{Zhumin Chu}
\affiliation{%
  \institution{Tsinghua University}
  \streetaddress{30 Shuangqing Rd}
  \city{Haidian Qu}
  \state{Beijing Shi}
  \country{China}}
 \email{chuzm19@mails.tsinghua.edu.cn}

\author{Qingyao Ai}
\authornote{Corresponding author.}
\affiliation{%
  \institution{Tsinghua University}
  \streetaddress{30 Shuangqing Rd}
  \city{Haidian Qu}
  \state{Beijing Shi}
  \country{China}}
 \email{aiqy@tsinghua.edu.cn}
 
\author{Zhihong Wang}
\affiliation{%
  \institution{Tsinghua University}
  \streetaddress{30 Shuangqing Rd}
  \city{Haidian Qu}
  \state{Beijing Shi}
  \country{China}}
  \email{wangzhh629@tsinghua.edu.cn}

\author{Yiqun Liu}
\affiliation{%
  \institution{Tsinghua University}
  \streetaddress{30 Shuangqing Rd}
  \city{Haidian Qu}
  \state{Beijing Shi}
  \country{China}}
  \email{yiqunliu@tsinghua.edu.cn}
  
\author{Yingye Huang}
\affiliation{%
  \institution{Tsinghua University}
  \streetaddress{30 Shuangqing Rd}
  \city{Haidian Qu}
  \state{Beijing Shi}
  \country{China}}
  \email{yy-huang19@mails.tsinghua.edu.cn}
 
\author{Rui Zhang}
\affiliation{%
  \institution{Tsinghua University}
  \state{Beijing Shi}
  \country{China}}
  \email{rayteam@yeah.net}

\author{Min Zhang}
\affiliation{%
  \institution{Tsinghua University}
  \streetaddress{30 Shuangqing Rd}
  \city{Haidian Qu}
  \state{Beijing Shi}
  \country{China}}
  \email{z-m@tsinghua.edu.cn}

\author{Shaoping Ma}
\affiliation{%
  \institution{Tsinghua University}
  \streetaddress{30 Shuangqing Rd}
  \city{Haidian Qu}
  \state{Beijing Shi}
  \country{China}}
  \email{msp@tsinghua.edu.cn}


\renewcommand{\shortauthors}{Chu et al.}

\begin{abstract}
With the increasing popularity of conversational search, how to evaluate the performance of conversational search systems has become an important question in the IR community. Existing works on conversational search evaluation can mainly be categorized into two streams: (1) constructing metrics based on semantic
similarity (e.g. BLUE, METEOR and BERTScore), or (2) directly evaluating the response ranking performance of the system using traditional search methods (e.g. nDCG, RBP and nERR). However, these methods either ignore the information need of the user or ignore the mixed-initiative property of conversational search. This raises the question of how to accurately model user satisfaction in conversational search scenarios. Since explicitly asking users to provide satisfaction feedback is difficult, traditional IR studies often rely on the Cranfield paradigm (i.e., third-party annotation) and user behavior modeling to estimate user satisfaction in search. However, the feasibility and effectiveness of these two approaches have not been fully explored in conversational search. In this paper, we dive into the evaluation of conversational search from the perspective of user satisfaction. We build a novel conversational search experimental platform and construct a Chinese open-domain conversational search behavior dataset containing rich annotations and search behavior data. We also collect third-party satisfaction annotation at the session-level and turn-level, to investigate the feasibility of the Cranfield paradigm in the conversational search scenario. Experimental results show both some consistency and considerable differences between the user satisfaction annotations and third-party annotations. In addition to user studies, we propose dialog continuation or ending behavior models (DCEBM) to capture session-level user satisfaction based on turn-level information. Our methods explicitly model the continuation and ending behaviors of users and significantly outperform the baselines under the third-party turn-level satisfaction annotation setting. This indicates the potential of user behavior modeling in conversational search evaluation.
\end{abstract}


\maketitle

\section{Introduction}
\label{sec:introduction}
Conversational search (ConvS)~\cite{gao2018neural, anand2020conversational, deriu2021survey, czyzewski2020agent} has attracted increasing attention in both information retrieval (IR) and natural language processing (NLP) communities. Compared to the traditional query-SERP (Search Engine Result Page) search paradigm, conversational search introduces the mixed-initiative property~\cite{radlinski2017theoretical, aliannejadi2021analysing} , i.e. both the user and the system can lead the search process through conversations. Users can initiate, revise or change their information needs at any time, while the system can ask clarifying questions~\cite{aliannejadi2019asking} to better understand the user's information needs. The ConvS paradigm helps the system understand the user's information needs more thoroughly and precisely, and thus produces better results for the user's search intent.

As the popularity of ConvS grows, how to evaluate the performance of a ConvS has become an important question to the IR community. Batch evaluation plays an indispensable role in the development and research of traditional IR systems~\cite{saracevic1995evaluation}. On the one hand, evaluation contributes to performance comparisons between models, enabling researchers to more explicitly identify which models are performing
better. On the other hand, IR models cannot be trained without the guidance of evaluation metrics. Some models even take a particular evaluation metric as the optimization target directly. Yet, the batch evaluation of ConvS, especially from the perspective of satisfying user's search intent, hasn't been fully explored in the literature. Specifically, existing conversational search evaluation works can be categorized into two streams:

The first way to evaluate ConvS is from the view of dialog systems, and to construct evaluation metrics (e.g. BLEU~\cite{papineni2002bleu}, METEOR~\cite{banerjee2005meteor} and BERTScore~\cite{zhang2019bertscore}) based on semantic similarity. Such methods first collect one or a set of ideal system responses, which are often generated by humans. The evaluation metrics are then computed by comparing the word overlap or semantic similarity between the system-generated responses and the ideal candidates. As a result, this evaluation paradigm naturally suffers from the following two shortcomings: 
(1) The answers to questions in the open-domain usually are diverse. As it is prohibitive to collect all possible ideal responses, existing ConvS datasets only provide limited number of golden responses for testing, which makes the evaluation metrics biased and unreliable. 
(2) Evaluation solely based on the semantic correlations between the human-generated responses and system-generated responses ignores the information demand properties of the user. Whether the user's search intention has been satisfied or not is hardly covered in the metrics.

The second way to evaluate ConvS is to directly measure the response ranking performance of the system using traditional search methods (e.g. nDCG~\cite{jarvelin2002cumulated}, RBP~\cite{moffat2008rank} and nERR~\cite{chapelle2009expected}). While these metrics are easy to compute based on the final retrieved results of a ConvS, they ignore the mixed-initiative property of ConvS and simply treats it as a ranking system. As a result, evaluating ConvS with traditional search metrics cannot capture question effectiveness, cognitive cost, and many important factors that affect the overall user satisfaction of a ConvS. To the best of our knowledge, how the interactions between users and ConvS in search sessions affect the perceived satisfaction of users is mostly unknown.

Considering the shortcomings of previous evaluation approaches, many efforts start to focus on user satisfaction in recent years~\cite{fu2022evaluating, siro2022understanding, bodigutla2019multi}. As an essential approach for traditional IR evaluation, user satisfaction is a fundamental indicator for the quality of system responses in conversational search. However, it is difficult to explicitly obtain user satisfaction in practice. Currently, there mainly exist the following two approaches to model user satisfaction: (1) the Cranfield paradigm~\cite{cleverdon1967cranfield}, i.e. using third-party annotators to label user satisfaction under appropriate annotation guidelines. The basic hypothesis of this approach is that user satisfaction is perceivable by third-party personnel. However, the plausibility of this hypothesis has not been sufficiently verified. Fu et al.~\cite{fu2022evaluating} found that third-party annotation shares a low degree of agreement with user annotation, yet these two hold similar patterns. More research is needed to further investigate this issue. (2) utilizing user behavior in ConvS for user satisfaction modeling~\cite{bodigutla2019multi}. This type of approach relies on sufficient turn-level and session-level conversational behavior and annotation information, especially turn-level satisfaction annotation results. In the field of dialog systems, some studies have attempted to aggregate turn-level user satisfaction to model session-level user satisfaction~\cite{zhang2021dynaeval}. However, in the ConvS domain, existing datasets are still incapable of meeting this research requirement.

In this paper, we dive into the evaluation of ConvS from the perspective of user satisfaction. We build a novel ConvS experimental platform and collect a sufficient amount of conversations with rich annotations and search behavior data, including turn-level and session-level user satisfaction annotations. In addition, we collect turn-level and session-level third-party satisfaction annotation, to explore the feasibility of the Cranfield paradigm in ConvS. We also try to establish an association between turn-level satisfaction and session-level evaluation based on user behaviors. Based on the user and third-party satisfaction annotation results, we perform a comprehensive data analysis in an effort to answer the following two research questions:


\begin{itemize}
	\item \textbf{RQ1}: Can we evaluate the quality of ConvS in terms of user satisfaction from a third-party view? What degree of agreement can be achieved between third-party annotations and users' explicit feedback?
	\item \textbf{RQ2}: How session-level user satisfaction can be modeled based on turn-level annotations of user satisfaction? 
\end{itemize}

The contributions of this paper can be summarized as follows:
\begin{enumerate}
	\item We construct a Chinese open-domain conversational search behavior dataset, the first ConvS dataset to collect both dialog contents and corresponding agent search behaviors for real information needs generated by users. The dataset is made public for academic usage\footnote{https://github.com/chuzhumin98/ConvSearch-Dataset}.
	\item We propose a framework to conduct third-party annotations for both turn-level and session-level satisfaction. The annotators are required to conduct multiple rounds of revision and consultation to reach a consensus on the annotation. We conduct several rounds of third-party satisfaction annotation experiments based on this framework.
	\item We analyze the consistency of both intra- and inter-third-party assessors and users in terms of various dimensions including different levels (session-level and turn-level), different annotation stages, and different grade reduction strategies. The cross-analysis presents a series of interesting findings.
	\item We validate the feasibility of constructing session-level satisfaction from turn-level information. We also propose novel dialog continuation and ending behavior models (DCEBM) to model the user's behaviors of continuing or ending a dialog. Experimental results show that the DCEBM models outperform all the statistical models based on session-level and turn-level information.
\end{enumerate}

\section{Related Work}
\label{sec:related work}

\subsection{User satisfaction related studies on ConvS}
User satisfaction, as the most essential way to reflect the performance of ConvS systems, has gradually gained more and more attention from researchers in recent years. Bodigutla et al.~\cite{bodigutla2019multi} created a new Response Quality annotation scheme by determining ratings for just the given turn rather than all preceding turns. Their experimental results showed a significantly high correlation between Response Quality ratings and explicit turn-level user ratings. Siro et al.~\cite{siro2022understanding} separated and then annotated satisfaction into six different aspects: relevance, interestingness, understanding, task completion, efficiency, and interest arousal under both turn-level and session-level. Their annotation results showed that the concept of satisfaction varies across different dialogs and assessors. Fu et al.~\cite{fu2022evaluating} performed ConvS tasks based on the SQuAD dataset~\cite{rajpurkar2016squad} and conducted satisfaction annotation by the user himself (direct) and crowdsourcing worker (indirect). They found a weak agreement of user satisfaction between direct and indirect assessments, but these two assessments share similar patterns.

\subsection{ConvS-related evaluation metrics}
Several of the existing evaluation metrics for conversational search derive ideas from question answering and dialog system approaches~\cite{liu2021meta}. They construct an ideal set of responses (ground truth) and evaluate the system by comparing the semantic similarity between the system-generated responses and the ideal responses. Typical metrics include BLEU~\cite{papineni2002bleu} and METEOR~\cite{banerjee2005meteor}. The main idea of these metrics is to calculate the overlap ratio of the $k$-gram elements between the candidate responses and ground truth. Such metrics of exact word match cannot capture the semantic similarity that may exist between different words. With the growing research in the field of word2vec, several word-embedding based semantic similarity metrics have been proposed, such as BERTScore~\cite{zhang2019bertscore} and Vector Extrema~\cite{forgues2014bootstrapping}. Such semantic similarity-based methods rely on the construction of ground truth. However, in open-domain ConvS, good responses are open and multivarie, and there does not exist a standardized answer. In addition, these semantic-based evaluation approaches only focus on evaluating relevance rather than characterizing the fulfillment of user information needs.

Another stream of research adopts traditional search methods (such as nDCG~\cite{jarvelin2002cumulated}, RBP~\cite{moffat2008rank} and nERR~\cite{chapelle2009expected}) to evaluate the document ranking performance of the ConvS systems. This research idea is close to the session search, but differs from the mainstream thinking about the shape of the ConvS system, thus limiting its usage scenario.

The PARADISE framework~\cite{deriu2021survey} inspires a category of dataset-specific metrics. The key idea of PARADISE is to adopt a trainable model to fit the manual judgments. Lowe et al.~\cite{lowe2017towards} proposed ADEM based on the PARADISE framework. ADEM adopts a recurrent neural network to predict the adequacy of each system response turn. It is worth noting that such learning-based metrics are strongly dataset-specific, thus limiting their generalizability.

\subsection{ConvS-related datasets}

In general, existing ConvS datasets can be divided into three main categories: algorithm evaluation-oriented, Wizard-of-Oz-based, and dialog-like platform-based. Next, we will introduce typical representatives of these three types of datasets.

\subsubsection{Algorithm Evalution-Oriented Datasets}
\label{subsec:cast}
With the purpose of evaluating conversational search systems, many datasets are designed with an emphasis on their reusability and comparability, such as CAsT-19/20/21~\cite{dalton2020cast} and MS MARCO datasets~\cite{nguyen2016ms}. In CAsT-19/20/21 datasets, the following questions generated by users only depend on the previous user turns rather than system responses. In the MS MARCO dataset, the organizers extract session samples from a commercial search engine, and then adopt the query sequence of each session as one dialog record.
Thus, in these settings, the generated response in the current turn would not influence the next one, of which the relevance can be annotated independently and used to evaluate systems. The evaluation could be problematic since the datasets are not produced interactively and the continuity of adjacent turns could be low. 

\subsubsection{Wizard-of-Oz-based Datasets}
\label{subsec:wizard-of-oz}
Since existing dialog systems can not yet generate high-quality dialog data competently, human-to-human~\cite{vtyurina2017exploring}, especially Wizard-of-Oz paradigm~\cite{kelley1984iterative} has attracted greater attention in recent years. In the Wizard-of-Oz paradigm, two participants are involved in the dialog. One plays the role of a user who addresses his or her specific information needs by engaging in a dialog with a wizard agent.  The other plays the role of a wizard agent who imitates to be an intelligent system and answers the user's questions by seeking relevant information. 

The typical Wizard-of-Oz datasets contain Wizard of Wikipedia~\cite{dinan2018wizard}, MultiWOZ~\cite{budzianowski2018multiwoz}, Topical-Chat~\cite{gopalakrishnan2019topical} and WISE~\cite{ren2021wizard}. However, none of these datasets satisfies the \textit{authenticity} property since their user information needs are assigned directly to users by organizers. In Wizard of Wikipedia, organizers collect conversation requests via a crowdsourcing platform. Task-oriented conversational information need is defined based on an ontology in MultiWOZ. While in Topical-Chat, the selected Washington Post articles and shortened Wikipedia lead sections about the top entities are directly assigned to both user and agent, prompting them to engage in a series of chit-chats with interchangeable identities. The WISE dataset, as the most similar Wizard-of-Oz dataset to ours, also contains agent search behaviors. However, their user information needs are distilled from the search logs of a commercial search engine and assigned to the users, rather than being naturally generated by themselves.

\subsubsection{Dialog-like Platform-based Datasets}
\label{subsec:msdialog}
Another approach to dataset construction is to crawl data from existing conversation-like platforms and reorganize them as a conversational search dataset. MSDialog dataset~\cite{qu2018analyzing} is a typical example of this type. Although this data acquisition method ensures the \textit{authenticity} of the data, the lack of agent search behavior discourages the widespread use in the research of system decision-making.

\section{Data Collection}
\label{sec:data collection}

Since the existing ConvS systems cannot be competent in providing high-quality conversational information search service as the agent, we adopt the widely used Wizard-of-Oz paradigm~\cite{vtyurina2017exploring} and filed study method in such situations to collect the high-quality dataset. Following the Wizard-of-Oz field study, we further conduct annotation experiments to enrich the dataset. In this section, we will introduce collected \textbf{C}hinese \textbf{O}pen-Domai\textbf{n} Con\textbf{v}ersational \textbf{Search} Behavior Dataset (ConvSearch).

\subsection{Weakness of existing ConvS datasets}
\label{subsec:weakness of existing convs datasets}
As a new search paradigm, ConvS remains largely unexplored, including user modeling~\cite{choi2019offline, lipani2021doing, azzopardi2018conceptualizing, trippas2018informing} and system construction~\cite{gao2018neural, sun2018conversational}. Since ConvS is highly interactive and proactive, modeling user intent and satisfaction~\cite{choi2019offline, bodigutla2019multi} is even more important for system design and evaluation than the traditional web search. Thus, a high-quality dataset with rich information collected in a real-life scenario would be greatly beneficial for the ConvS research community. Unfortunately, existing ConvS datasets often have two limitations that prevent them from serving the needs of this study:

\begin{enumerate}
	\item \textbf{Authenticity}: The information need in many datasets is predefined by the researchers rather than generated in-situ by users themselves. Thus, the realisticness of these datasets is hard to guarantee. 
	\item \textbf{Completeness}: Different from other intelligent assistance, like chatbot or customer service, seeking information is the key component in ConvS. However, information-seeking behaviors of agents are greatly ignored in most existing datasets, which only contain dialog contents. 
\end{enumerate}

Table~\ref{tab:comparison of convs datasets} shows the fulfillment of the two characteristics in the existing datasets and our ConvSearch Dataset. To the best of our knowledge, ConvSearch is the first dataset that satisfies both properties of \textit{authenticity} and \textit{completeness}.

\begin{table}[]
	\begin{center}
		\caption{Comparison of related conversational search datasets}
		\label{tab:comparison of convs datasets}
		\vspace{-3mm}
		\begin{tabular}{lcc}
			\toprule[1.0pt]	
			Dataset & Authenticity  & Completeness  \\	
			\midrule[0.6pt]
			CAsT-19/20/21~\cite{dalton2020cast} &  &  \\	
			MS MACRO~\cite{nguyen2016ms} & \checkmark~\tablefootnote{The MS MACRO only contains the real-life user questions, but not the system response.} & \\
			MultiWOZ~\cite{budzianowski2018multiwoz} &  &  \\
			Topical-Chat~\cite{gopalakrishnan2019topical} &  &  \\
			MSDialog~\cite{qu2018analyzing} & \checkmark &  \\
			WISE~\cite{ren2021wizard} &  & \checkmark \\
			ConvSearch (ours) & \checkmark & \checkmark \\
			\bottomrule[1.0pt]
		\end{tabular}
	\end{center}
	\vspace{-3mm}
\end{table}

\subsection{Experimental platform}
\label{subsec:experimental platform}
To better collect conversational search data, we develop a novel conversational search experimental platform. The platform is implemented based on \textit{Django} framework\footnote{https://www.djangoproject.com/}. Figure~\ref{fig:overall framework for data collection} shows the overall framework of data collection using this experimental platform. The following shows its main functional modules:

\begin{figure}[thbp]
    \captionsetup[subfigure]{justification=centering}
   \centering
   \includegraphics[width=0.8\textwidth]{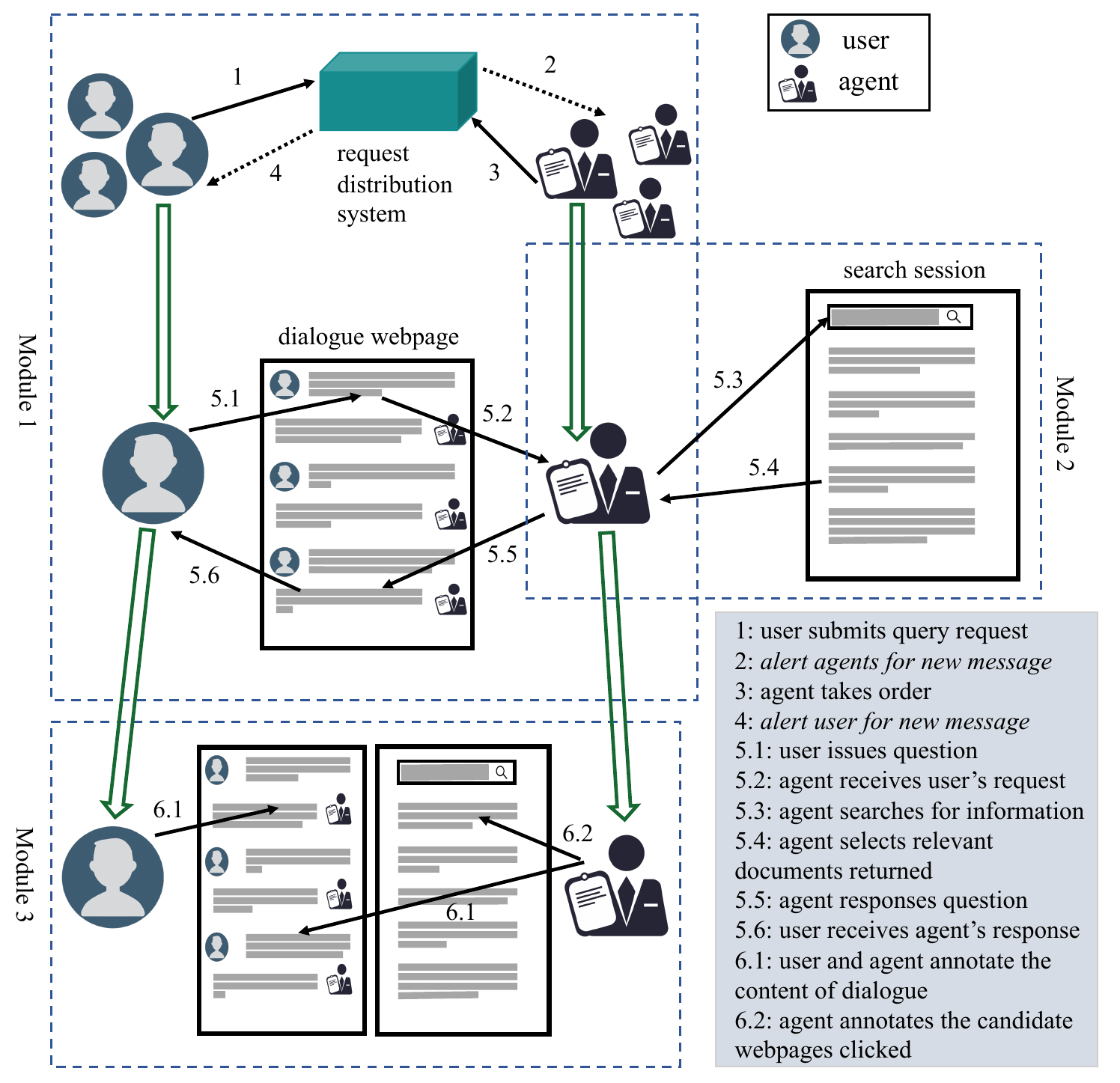}
   \caption{The overall framework for data collection.}
   \label{fig:overall framework for data collection}
 \end{figure}

\subsubsection{Support the user-agent dialogs}
Since the agent is played by the human which cannot wait for the request from users at any time, in reality, we adopt the asynchronous request distribution strategy. Specifically, inspired by the popular takeaway delivery platform we develop a message reminder module, which can broadcast all agents when one or more users submit requests and any agent who is available at that time can actively choose one from the request pool. By this way, we can to some extent guarantee that the user requests can be answered in timely and the dialog process will not be interrupted accidentally, thus the collected dataset quality is improved.

\begin{figure}[thbp]
    \captionsetup[subfigure]{justification=centering}
   \centering
   \includegraphics[width=0.45\textwidth]{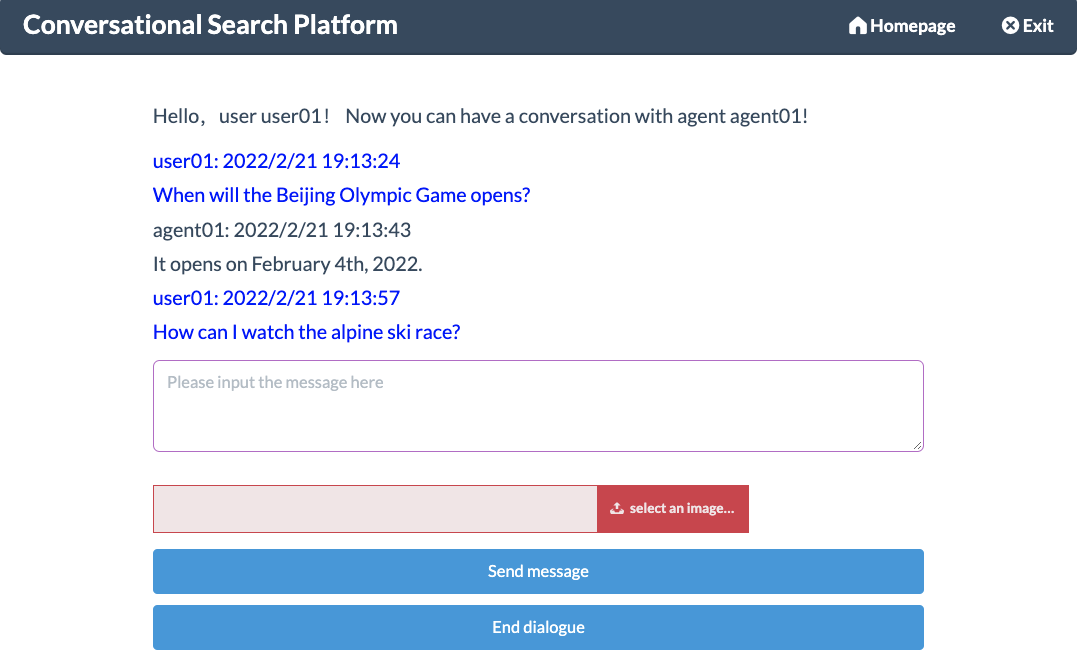}
   \caption{Experimental platform interface of \textit{dialog} webpage from user's view.}
   \label{fig:dialog webpage user}
 \end{figure}
 
 After the user-agent entering into the dialog stage, they can start communicating using the message sender box in text, link or image format on the dialog webpage. The dialog history is presented on this webpage. Since it is only the user instead of the agent that can end the conversation during the dialog, an end-dialog button is placed at the bottom of the user-side, as shown in Figure~\ref{fig:dialog webpage user}. 


\subsubsection{Support for agents to seek information in response to user requests}
\label{subsec:support for agents to seek information in response to user requests}
To collect the search behavior data of agents, every agent is required to use external source to acquire information needed to response to the user. Thus, several popular Chinese search engines and knowledge acquisition platforms are integrated into the agent-side dialog webpage. As such, agents can easily choose one or more sources to access the information whenever they need. 

We also develop a Chrome extension\footnote{https://developer.chrome.com/docs/extensions/mv3/} to record agent search behavior, including query (re)formulation, webpage browsing, click action, et al. With the permission of the agent, the Chrome extension is installed successfully on the agents' computer before the field study starts. 

\subsubsection{Support for user-agents to carry out post-dialog annotations}
Soon after each dialog, both user and agent need to visit the \textit{annotation} webpage to assess the dialog contents. The agent is required to provide responses to the user based on the acquired information from external sources instead of his or her own knowledge base. The turn is presented and annotated one after another. The dialog as the whole is annotated when the annotation of all the turns are completed. The specific annotation items as well as their annotation rules could be found in our dataset homepage~\footnote{https://github.com/chuzhumin98/ConvSearch-Dataset/blob/main/annotation-rules/Participant\_Annotation\_Rules.pdf}.

\subsection{Field Study}
\label{subsec:field study}

We recruit 25 agents (44\% male and 56\% female) and 51 users (33\% male and 67\% female) to participate in the field study. These participants are students or staff in a university. All participants are aged between 18 and 35 years old, except for one 53-year-old staff who acts as a user.

Before the field study starts, we hold a training session for agents and users separately face-to-face. The training items contain the field study purposes and procedures, as well as the risks, benefits and privacy concerning the study. All participants sign the informed consent. 

During the training process, all participants are instructed on the Wizard-of-Oz setup as well as annotation rules, and are made familiar with the experimental platform. We also compile a detailed guide for further reference by participants. For agents, we ask them to imitate the AI system when seeking information during the conversation. In addition, to perform the general information-seeking task usually progressing from submitting the query to selecting and assessing the retrieved result, the agents also need to either extract the information or generate a response based on the selected documents. 

The field study lasts 45 days from November 24th, 2021, to January 8th, 2022. . Before the large-scale field study, we recruit 10 users and agents to conduct a small-scale pilot experiment to ensure that the experimental design, platform function and training process are problem-free. In total, we collect 1,131 valid dialogs after removing politically sensitive, duplicative, low-quality and incomplete dialogs. The total cost of the field study is about 8,000 dollars (including user intent and agent action annotation), with an average of $8$ dollars per dialog.
Table~\ref{tab:basic statistics of convsearch dataset} shows the basic statistics of the ConvSearch dataset. More detailed statistics and analysis can be found on the homepage.

\begin{table}[t]
\vspace{-3mm}
	\begin{center}
		\caption{Basic statistics of ConvSearch dataset}
		\label{tab:basic statistics of convsearch dataset}
		\vspace{-3mm}
		\begin{tabular}{lr}
			\toprule[1.0pt]	
			Statistics & Number  \\	
			\midrule[0.6pt]
			\# Dialogs & 1,131 \\
			Average total turns per dialog & 8.30 \\
			Average merged turns per dialog & 6.89 \\
			Average agent queries per dialog & 6.01 \\
			Average agent queries with clicks per dialog & 3.68 \\
			Average terms per agent query & 10.37 \\
			Average agent landing pages per dialog & 8.78 \\
			\bottomrule[1.0pt]
		\end{tabular}
	\end{center}
	\vspace{-3mm}
\end{table}

\subsection{User Intent and Agent Action Annotation}
\label{subsec:user intent and agent action annotation}
Inspired by ~\cite{qu2018analyzing, ren2021wizard}, we group user intent and agent action into multiple categories. Table~\ref{tab:categorization rules and examples for user intent} and Table~\ref{tab:categorization rules and examples for agent action} show the specific annotation rules for user intent and agent action, respectively. Since \textit{reveal} in user intent and \textit{answer} in agent action occupies a large proportion of the dataset, we divide these two categories into more fine-grained sub-categories.

Due to the complexity of the conversation content, the assessors are allowed to annotate one turn with two labels\footnote{Single label cases are also allowed to occur.}: a primary label and a secondary label. A primary label and secondary label should be selected according to the major and minor contribution of the current turn content to the whole dialog. For example, the user says ``\textit{Jane Eyre sounds interesting! How can I buy it?}" (the last response of agent is ``\textit{I recommend Jane Eyre and A Tale of Two Cities}"). The first sentence ``\textit{Jane Eyre sounds interesting!}" is feedback towards \textit{Jane Eyre}, so it should be labeled as \textit{others}. ``\textit{How can I buy it?}" expresses an information need directly inspired by the agent's last response, thus it belongs to the \textit{reveal-inspire} label. The request in the latter sentence is further answered by the agent, making more contributions in motivating the following dialog. Thus, the appropriate annotation of this example is \textit{reveal-inspire} (primary) + \textit{others} (secondary). 

In addition, the \textit{answer} category in agent action involves more complex two-class scenarios:

\begin{itemize}
	\item If a part of the agent response can already satisfy the user's information need completely, then the category of this part is the primary label, while the category that the rest content belongs to is the secondary label.
	\item If the user's information need contains multiple questions, then the category of the longer response to one of the questions is the primary label, and the category of the shorter is the secondary label.
\end{itemize}

\begin{table*}[t]
\small
\begin{center}
\caption{Categorization rules and examples for user intent}
\label{tab:categorization rules and examples for user intent}
\begin{tabular}{llm{4cm}m{5cm}m{4cm}}
\toprule[1.0pt]	
\multicolumn{2}{l}{Intent}         & Description & Example & Remark  \\
\midrule[0.6pt]

\multirow{3}{*}{reveal} & initiate &   Express information need. Start a new dialog topic.          &   
\makecell[l]{(dialog starts)\\User: Can you recommend a book? \\ (reveal-initiate) \\ 
Agent: I recommend ... \\ User: When will the Beijing Olympic \\
 Game opens? (reveal-initiate)}
   &  First question, or question with the information need that is not related to the previous conversation topic.    \\ \cline{2-5}
   
     & continue &   Express information need. Continue the topic user has mentioned above.  &   
\makecell[l]{User: When will the Beijing Olympic \\
 Game opens? \\
 Agent: It opens on February 4th, 2022.\\
 User: How can I watch the alpine ski\\ race? (reveal-continue) }   &  The information need is relevant to the previous dialog topic and not directly inspired by prior agent responses.   \\ \cline{2-5}
     
    & inspire  &  Express information need. The intent is directly inspired by prior agent responses.        & 
\makecell[l]{User: Can you recommend a book? \\
Agent: I recommend \textit{Jane Eyre}. \\
User: How can I buy it? (reveal-inspire)}    &  The information need must be inspired directly from prior agent responses, with no interruptions from other user questions that utilizes the inspired information.    \\ \hline

revise    &     &  Revise the intent proactively when the expression of previous information need is wrong or unclear, e.g., typos, grammatical errors or unclear expression.           &  
\makecell[l]{User: When will the Olympic Game\\ opens? \\
 User: I mean Beijing Winter Olympic \\Game. (revise)}     & User proactively revises the information need, rather than  interpreting the need passively after agent asks clarifying question.   \\ \hline
     
interpret  &    & Interpret or refine the intent by answering the clarification question from the agent.    &  
\makecell[l]{User: When will the Olympic Game\\ opens? \\
Agent: You mean Beijing Winter\\ Olympic Game?\\
User: Yes. (interpret)}    &  The interpretation is made after agent asks clarifying question.   \\ \hline

chitchat  &    &  Greetings or other contents that are unrelated to the information need.           &   
\makecell[l]{User: Thanks for your help. (chitchat) \\
User: Are you there? (chitchat)}   &     \\ \hline

others   &   & Other user intent that are not included in the above mentioned categories.             &  
\makecell[l]{User: This movie sounds interesting!\\ (others) \\
User: I got it. (others)}   &     \\ 
\bottomrule[1.0pt]
\end{tabular}
\end{center}
\end{table*}

 \begin{table*}[t]
 \small
\begin{center}
\caption{Categorization rules and examples for agent action}
\label{tab:categorization rules and examples for agent action}
\begin{tabular}{llm{3.5cm}m{4.5cm}m{5cm}}
\toprule[1.0pt]	
\multicolumn{2}{l}{Intent}         & Description & Example & Remark   \\
\midrule[0.6pt]
clarify    &     &  Ask questions to clarify user intent when the intent is unclear or exploratory.           & 
\makecell[l]{User: I want to know something \\about sun. \\
Agent: Do you mean stellar sun \\or company Sun? (clarify)}     &  The clarifying question could be yes-or-no question, choice question or even open question.    \\ \hline

\multirow{3}{*}{answer} & single-fact &   Give a unique and unambiguous fact. The answer is objective and certain.          & 
\makecell[l]{Agent: The price of this ceil phone\\ is 400 dollars. (answer-single-fact) \\
Agent: The three primary colors\\ of light are red, green and blue.\\ (answer-single-fact)}    & Single-fact does not mean that there is only one entity in the answer, but instead that there is only one answer in the correct answer space.     \\ \cline{2-5}

   & multi-fact &  Give some elements/aspects or entities of facts. The total answer fact space is certain and large. Some of the facts can satisfy the user's information need.           & 
\makecell[l]{Agent: Einstein's life experiences\\ are as follows: ...... \\(answer-multi-fact) \\
Agent: One of the reservation \\numbers is xxx. \\(answer-multi-fact)}     & In contrast to single-fact, multi-fact means that there exist multiple answers in the correct answer space. Either some or all of the answers can satisfy the user's information need.    \\ \cline{2-5}
   
   & opinion  &  Give instructions, advice, or ideas. The answer can come from personal opinion of the agent or some netizens, or it has been recognized as a social consensus.   &
\makecell[l]{Agent: To diminish wrinkles,\\ I recommend that you can ...... \\(answer-opinion) \\
Agent: The historical meaning\\ of the Renaissance Movement is\\ as follows: ...... (answer-opinion)}        &  If the correctness of answer varies with different people's perceptions, the answer belongs to \textit{opinion}, otherwise it is a \textit{fact}.  \\ \hline
   
no-answer    &     &  Notice to the user that the relevant information has not been found           &   Agent: Sorry, no product has been found to satisfy your requirements. (no-answer)   &  Notice the difference between \textit{I have not found any products that meet the requirements} (no-answer) and \textit{Apple have not marketed a product that meets the requirements} (answer-single-fact). (User question is \textit{please recommend an iPhone with 2,048GB disk space}.)  \\ \hline

chitchat  &    & Greetings or other contents that are unrelated to the information need.            &  
\makecell[l]{Agent: My pleasure to help you. \\(chitchat) \\
Agent: Hello! (chitchat)}    &     \\ \hline

others   &   & Other agent actions that are not included in the above mentioned categories.            &  Agent: Any more questions do you have? (others)    &    \\ 
\bottomrule[1.0pt]
\end{tabular}
\end{center}
\end{table*}
 
To ensure the annotation quality, we develop a detailed and understandable instruction with examples and a screening mechanism to recruit higher quality annotators. we initially recruit 15 assessors and conduct a training program for each individual lasting about 1 hour.  After training, all assessors enter into the trial stage where they need to annotate 50 to 150 dialogs. Then, one of the authors examines these trial annotation results and select 6 assessors among 15. According to the common errors made at the trial stage, we organize another training program to further explain the annotation rules for these 6 assessors. We also provide a more detailed guideline with examples for them. These 6 assessors are divided into 2 groups. Each group of assessors needs to complete the assigned annotation task with about 550 dialogs, half of the dataset, such that each dialog is annotated by 3 different annotators.

\subsection{Third-party user satisfaction annotation experiment}
\label{sec:Third-party user satisfaction annotation experiment}

\begin{figure*}[thbp]
    \captionsetup[subfigure]{justification=centering}
   \centering
   \includegraphics[width=0.9\textwidth]{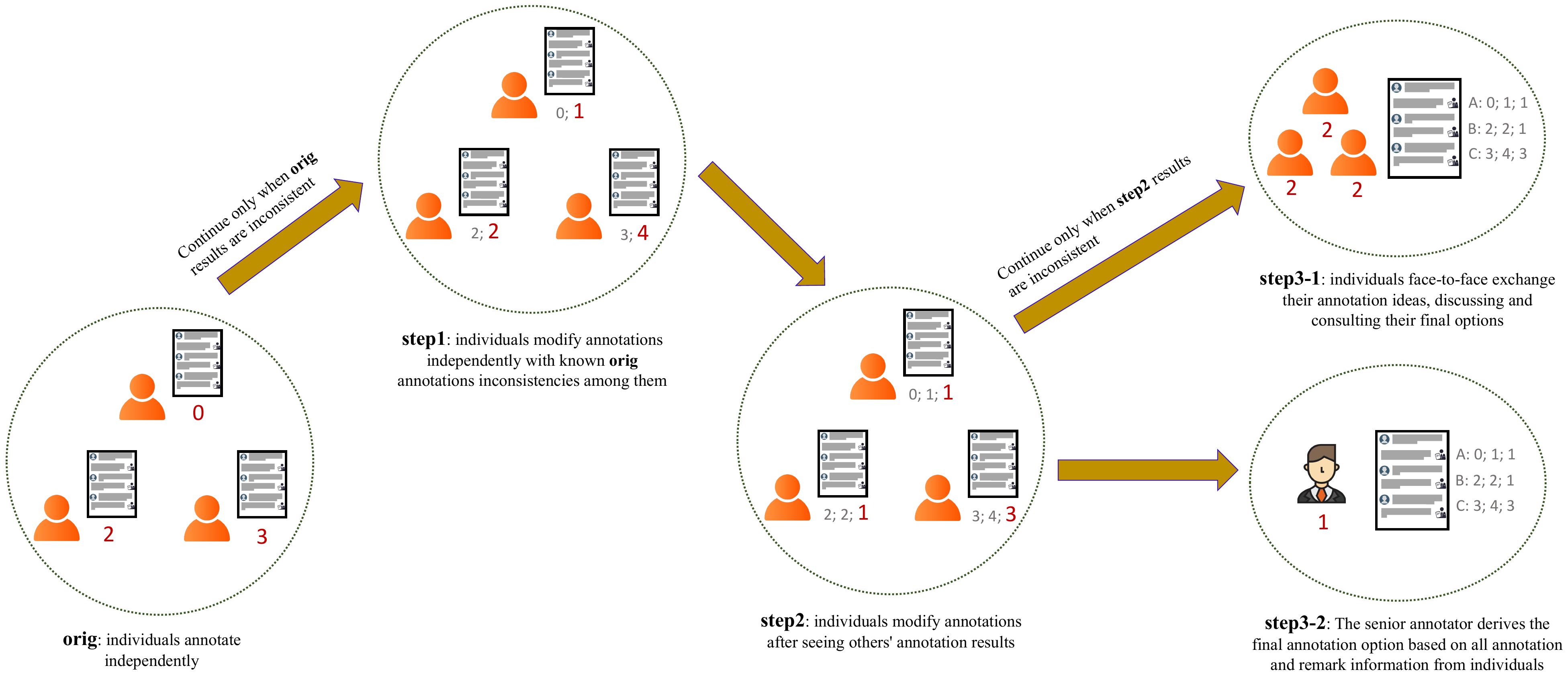}
   \caption{Experiment procedures for the third-party user satisfaction annotations in Sec.~\ref{sec:Third-party user satisfaction annotation experiment}}
   \label{fig:experiment procedures for the third-party user satisfaction annotations}
 \end{figure*}
 
 To investigate to what extent third-party annotations can be used to replace users' satisfaction annotations, we conduct a third-party annotation experiment with multiple steps and various strategies. Fig~\ref{fig:experiment procedures for the third-party user satisfaction annotations} summarizes the whole procedure of the third-party annotation experiments.
 
 We recruit $9$ third-party annotators, grouped by three of each, to perform the follow-up third-party annotation experiment. 
 
 \subsubsection{\textit{orig}}
 At the beginning of the experiment, we train the annotators and quickly familiarize them with the labeling specification and process through $5$ trial annotation tasks. We ask the assessors to put themselves into the role of user as much as possible and perform turn-level and session-level user satisfaction annotations from the user's perspective. In order to mimic the scenario of post-dialog satisfaction annotation by the user, the self-developed annotation system presents the entire dialog to the annotator in prior to the annotation of each dialog. Afterwards, the annotator is required to assess the turn-level satisfaction one turn after another in the order of their appearances in the dialog. After the completion of all turns, the annotator is asked to evaluate the session-level user satisfaction.


Considering high inconsistency among the original third-party annotation results (more details in Sec.~\ref{subsec:consistency within third-party annotation}), we further conduct three rounds of modification experiment, as Fig~\ref{fig:experiment procedures for the third-party user satisfaction annotations} shows.

\subsubsection{\textit{step1} and \textit{step2}}
For those turns and dialogs that do not achieve agreement, we ask each assessor to perform two rounds of revision independently at first, as \textit{step1} and \textit{step2} in Fig~\ref{fig:experiment procedures for the third-party user satisfaction annotations}. At \textit{step1}, we notify each annotator that there is no agreement among the three annotators. Then, they shall review the dialog content carefully to decide whether to modify the corresponding turn-level and session-level satisfaction annotations. Then, at \textit{step2}, the \textit{orig} annotation results are presented to each annotator. The annotator needs to decide whether to modify his/her annotation result referring to the \textit{orig} annotation of the other two annotators. All annotators are required to write down the reasons for their decisions at \textit{step1} and \textit{step2}, which ensures that they are conscientious in the revision process of these two steps.

\subsubsection{\textit{step3-1} and \textit{step3-2}}
Although a much higher proportion of turns and dialogs have reached agreement on the satisfaction rating among annotators after \textit{step1} and \textit{step2}, there still exist many cases that cannot achieve consistency across annotators. In this regard, we propose two approaches to merge the opinions of annotators: 

The first approach (\textit{step3-1}) is to organize a consultation session gathering the three annotators. In the session, each annotator shares his or her perspective underlying his or her label, and ultimately a consensus will be achieved by consultation and discussion (in the case of a few cases, they are allowed to end up with different labeling results if they still cannot reach agreements even after consultation). Considering the high cost of this approach, we only conduct \textit{step3-1} in the more complicated situation, session-level satisfaction, so as to investigate the effectiveness of consultation sessions at an affordable cost.

Another approach (\textit{step3-2}) is inspired by the paper reviewing procedure, to introduce a more experienced "senior annotator" (one of the authors), needs through reasons provided by the three annotators in the previous steps, and derives the final merged third-party user satisfaction annotation results on the basis of respecting the opinions of all parties.

%

\section{Consistency within third-party annotation}
\label{subsec:consistency within third-party annotation}
In the following three sections, we conduct a series of experiments and analyses based on the ConvSearch dataset, trying to answer the two research questions proposed in Sec.~\ref{sec:introduction}.

To answer RQ1, we first analyze the agreement within third-party annotations. We use Fleiss' $\kappa$~\cite{fleiss1969large} to evaluate it, as Table~\ref{tab:intra third assessors agreement} shows. Table~\ref{tab:intra third assessors agreement} shows the annotation consistency under different annotation stages and different grade reduction strategies. The consistency of the original third-party annotation is quite low, with the Fleiss' $\kappa$ values at both the session-level and turn-level being only around $-0.1$. When we merge the annotation grades appropriately, there is a significant improvement in annotation consistency, especially after merging the grade $3$ and $4$. This phenomenon is particularly evident at the turn-level, where the improved $\kappa$ value is more than $0.3$. It indicates that grade $3$ (satisfactory) and grade $4$ (highly satisfactory) are quite confusing for third-party assessors, because different assessors might label a fairly high-quality agent response as $3$ instead of $4$ due to different subjective factors, such as lengthy responses, lack of artwork and etc. Table~\ref{tab:intra third assessors agreement} also shows that the consistency among third-party assessors continuously improves as the annotation modification stage advances. This is in line with our expectation that the goal of the modification and consultation process is to allow annotators to exchange annotation ideas with each other and form a unified annotation consensus.

\begin{table}[]
    \caption{The annotation agreement (Fleiss' $\kappa$) among three third-party assessors under different annotation stages and different grade reduction strategies. The second row of the table lists various reduction strategies. For example, \textit{(01)(2)(34)} denotes that the original annotation grades $0$ and $1$, $3$ and $4$ are combined respectively to obtain the reduced three-grade annotation results. \textit{Step3-2} is excluded from the table because it only contains the annotation results of one senior annotator.}
    \label{tab:intra third assessors agreement}
\begin{tabular}{c|cccc|cccc}
\hline
\multirow{2}{*}{stage} & \multicolumn{4}{c|}{session-level}                                    & \multicolumn{4}{c}{turn-level}                                       \\ \cline{2-9} 
                       & (0)(1)(2)(3)(4) & (0)(1)(2)(34)   & (01)(2)(34)     & (012)(34)       & (0)(1)(2)(3)(4) & (0)(1)(2)(34)   & (01)(2)(34)    & (012)(34)       \\ \hline
\textit{orig}          & -0.1344         & 0.1185          & 0.1182          & 0.1724          & -0.1252         & 0.1916          & 0.2096         & 0.2828          \\
\textit{step1}         & 0.1847          & 0.3339          & 0.3337          & 0.4176          & 0.0996          & 0.3054          & 0.3384         & 0.4147          \\
\textit{step2}         & 0.3981          & 0.4848          & 0.4847          & 0.5645          & \textbf{0.3666} & \textbf{0.4386} & \textbf{0.475} & \textbf{0.5485} \\
\textit{step3-1}       & \textbf{0.9608} & \textbf{0.9864} & \textbf{0.9864} & \textbf{0.9861} & -               & -               & -              & -               \\ \hline
\end{tabular}
\end{table}

\begin{figure}[thbp]
   \captionsetup[subfigure]{justification=centering}
   \centering
	\subfloat[session-level annotations]{
     \includegraphics[width=0.45\textwidth]{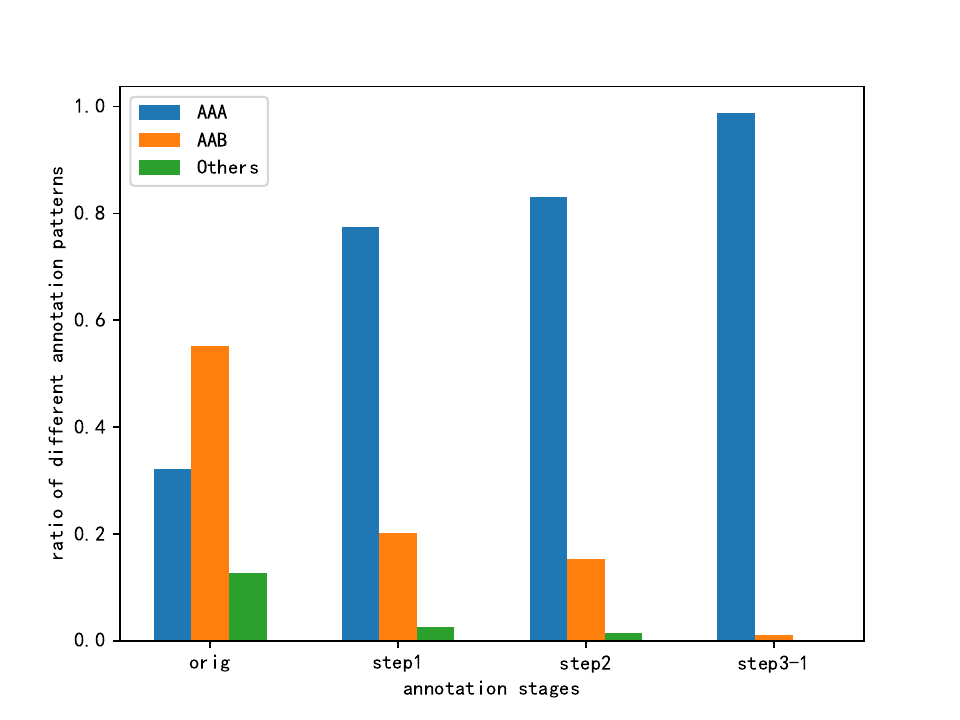}
     \label{fig:session-level annotations, ratio of pattern}
   }
   	\subfloat[turn-level annotations]{
     \includegraphics[width=0.45\textwidth]{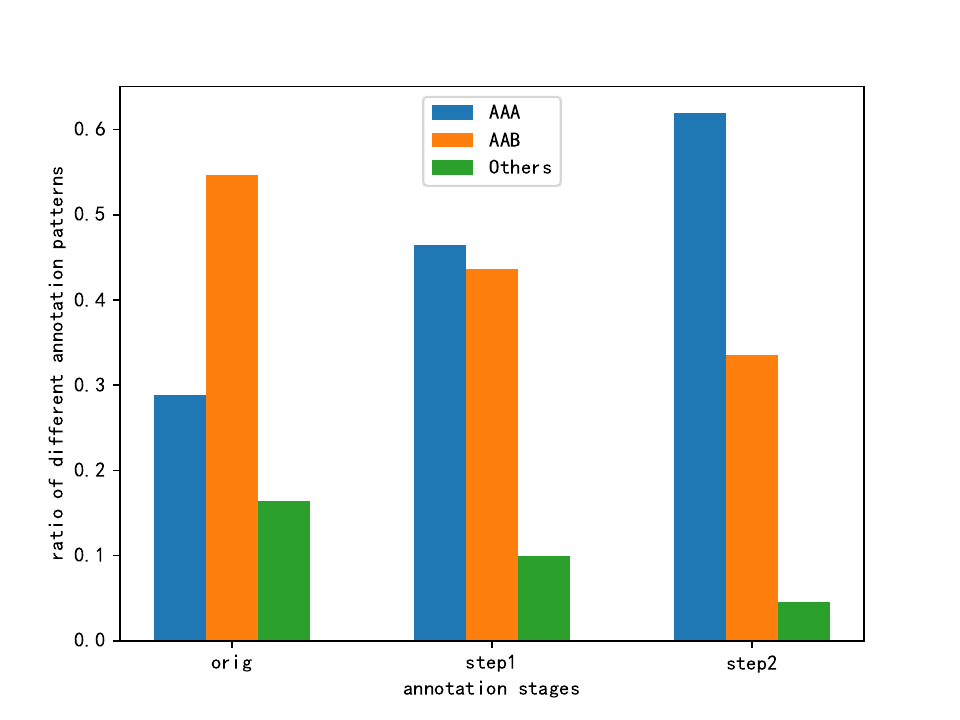}
     \label{fig:turn-level annotations, ratio of pattern}
   }
   \caption{The ratio of each annotation pattern among three third-party assessors in both session-level and turn-level satisfaction annotation}
   \label{fig:The ratio of each annotation pattern among three third-party assessors in both session-level and turn-level satisfaction annotation}
\end{figure}

To analyze the annotation consistency among third-party assessors more thoroughly, we measure the annotation patterns of each group of three annotators. Fig~\ref{fig:The ratio of each annotation pattern among three third-party assessors in both session-level and turn-level satisfaction annotation} shows the ratio of each annotation pattern at both session-level and turn-level satisfaction annotation. \textit{AAA} means that the three assessors annotate the same grade for the item (dialog or turn), e.g. 2-2-2 or 4-4-4. \textit{AAB} means that two assessors provide the same annotation, while another assessor differs from them with 1 grade, e.g. 2-2-3 or 3-4-4. \textit{Others} represents all other possible annotation patterns except \textit{AAA} and \textit{AAB}, such as 1-2-3 or 2-4-4 and etc. Fig~\ref{fig:The ratio of each annotation pattern among three third-party assessors in both session-level and turn-level satisfaction annotation} shows that more than $85\%$ of the original third-party annotations at both session-level and turn-level yield \textit{AAA} or \textit{AAB} annotation patterns, which indicates that different assessors are able to achieve a relatively consistent opinion on most items in the original annotations.

\begin{figure}[thbp]
   \captionsetup[subfigure]{justification=centering}
   \centering
	\subfloat[\textit{AAB+} pattern]{
     \includegraphics[width=0.45\textwidth]{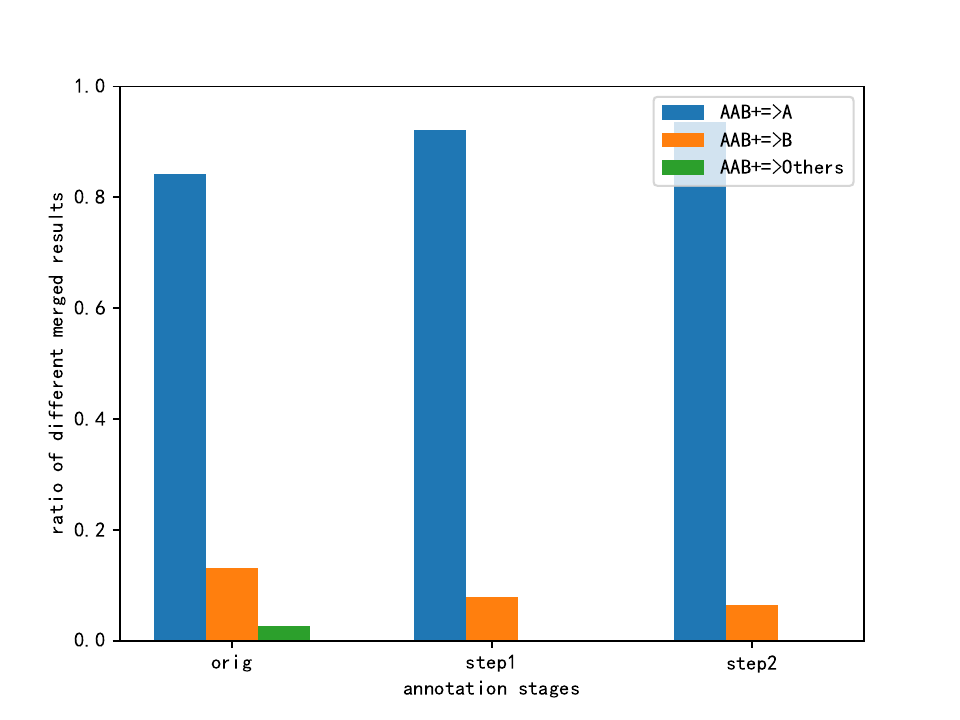}
     \label{fig:AAB+ pattern, ratios}
   }
   	\subfloat[\textit{AAB-} pattern]{
     \includegraphics[width=0.45\textwidth]{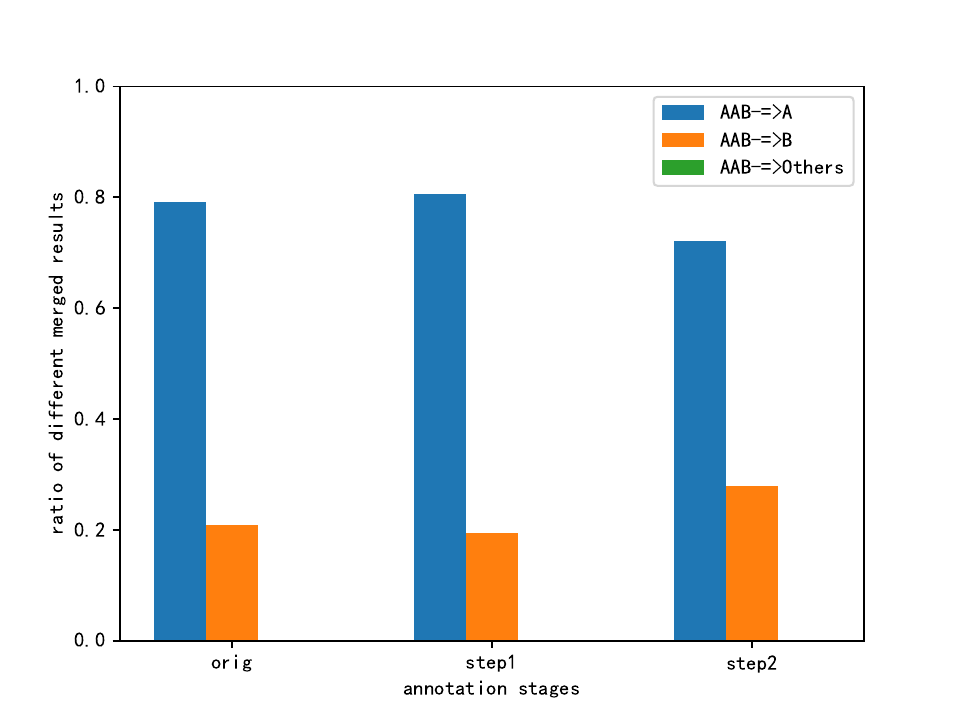}
     \label{fig:AAB- pattern, ratios}
   }
   \caption{The ratio of different types of third-party assessors merged results under \textit{AAB+} pattern and \textit{AAB-} pattern}
   \label{fig:The ratio of different types of third-party assessors merged results under AAB+ pattern and AAB- pattern}
\end{figure}

As we all know, majority voting is a frequently used method to merge the annotation results of multiple annotators. We compare the merged results by majority voting with the unified opinion of third-party annotators after discussion at \textit{step3-1}. \textit{AAB}, as the most frequent pattern in the data, is employed as the analytic object. More specifically, we divide \textit{AAB} into two subtypes, \textit{AAB+} and \textit{AAB-}. The difference between them lies in whether grade \textit{B} is higher (\textit{AAB+}, such as 2-2-3 and 3-3-4) or lower (\textit{AAB-}, such as 1-2-2 and 3-4-4) compared with the majority grade \textit{A}. Fig~\ref{fig:The ratio of different types of third-party assessors merged results under AAB+ pattern and AAB- pattern} shows the consistency ratio between majority voting results at each step and unified opinions at \textit{step3-1} in terms of  \textit{AAB+} pattern and \textit{AAB-} pattern. The results demonstrate the soundness of the majority voting strategy: The results of majority voting at all steps is highly consistent with the unified opinions of annotators after discussion. Fig~\ref{fig:The ratio of different types of third-party assessors merged results under AAB+ pattern and AAB- pattern} also demonstrates the difference between \textit{AAB+} and \textit{AAB-}: compared to \textit{AAB+}, the unified results of \textit{AAB-} are more skewed towards the minority opinion, and this tendency is further strengthened after one or two rounds of revision, at \textit{step1} and \textit{step2}. The opposite occurs with \textit{AAB+}, where the majority opinion of the assessors is increasingly skewed throughout the first two rounds of revisions. This phenomenon suggests that low-satisfaction annotations receive more attention from assessors in the revision and consultation process. This is probably because the assessors often rate lower satisfaction after capturing some signals in the dialog that make them feel unsatisfied. The two rounds of the revision process make this low-satisfaction belief even stronger. Thus, the lower-satisfied assessor still has a moderate chance to persuade the other two assessors during the consultation process.

To answer RQ1, we find that the original third-party annotation results yield moderate consistency, with more than $85\%$ of items following the \textit{AAA} or \textit{AAB} patterns. In line with our expectations, the consistency of the annotators increases as revision and consultation proceed. We also discover that assessors place greater importance on low-satisfaction views being the minority compared to high-satisfaction ratings as the minority.

\section{Consistency between third-party annotation and user annotation}
\label{subsec:consistency between third-party annotation and user annotation}

In this section, to better investigate the consistency between third-party annotation and user own annotation, we conduct analysis by both pointwise and pairwise comparison. Then we perform case studies on the dialogs with a large difference between these two kinds of annotations, and summarize the main reasons underlying the gaps.

\subsection{Pointwise analysis}
\label{subsubsec:Pointwise analysis}

We first explore the relationship between the third-party majority voting annotation results and the user's annotations. We focus on the dialogs where the original third-party annotation pattern is \textit{AAA} or \textit{AAX}. Here, \textit{AAX} means that two assessors yield the same annotation result while the other one differs from them, e.g., 2-2-4 and 3-4-4. Table~\ref{tab:AAA(AX) + AAX(AXY)} shows the distribution of the user's session-level and turn-level satisfaction under third-party \textit{AAA} and \textit{AAX} patterns with different reduction strategies. The numbers in Table~\ref{tab:AAA(AX) + AAX(AXY)} represent the proportion of particular user annotation under a certain third-party annotation pattern. For example, the value $0.3697$ in the third column of the first row indicates that there exists $36.97\%$ of the \textit{AAX}-pattern dialogs where users lean with the minority grade (X) on no grade reduction condition. A huge percentage boost in the consistency of user and third-party annotation can be observed when grades $3$ (satisfied) and $4$ (highly satisfied) are merged. This result demonstrates that grade $3$ and grade $4$ are likely to be confused by third-party assessors and users, since different individuals might have different measures on what kind of dialogs reach the high satisfaction grade. This finding coincides with the conclusion in Sec~\ref{subsec:consistency within third-party annotation}. Table~\ref{tab:AAA(AX) + AAX(AXY)} also shows that the perceptions of satisfaction between third-party assessors and users have considerable differences. Even after grade reduction, users are still aligned with the views of the minority of third-party labelers rather than the majority in about one-third of the conversations and turns. This phenomenon suggests that the traditional majority voting method may not be well suited to the ConvS scenario. The annotation opinions of the minority cannot be ignored when merging annotations.

\begin{table}[]
\caption{The distribution of user's own session-level and turn-level satisfaction under third-party \textit{AAA} and \textit{AAX} patterns on \textit{orig} stage with different reduction strategies. The capital letters in brackets denote the user's annotation results. \textit{A}, \textit{X} and \textit{Y} denote three different annotation grades.}
\label{tab:AAA(AX) + AAX(AXY)}
\begin{tabular}{l|cccc|cccc}
\hline
\multirow{2}{*}{reduction} & \multicolumn{4}{c|}{session-level} & \multicolumn{4}{c}{turn-level}    \\ \cline{2-9} 
                           & AAA(A) & AAX(A) & AAX(X) & AAX(Y) & AAA(A) & AAX(A) & AAX(X) & AAX(Y) \\ \hline
(0)(1)(2)(3)(4)            & 0.6171 & 0.4788 & 0.3697 & 0.1515 & 0.7069 & 0.5451 & 0.2984 & 0.1566 \\
(0)(1)(2)(34)              & 0.9075 & 0.6772 & 0.2169 & 0.1058 & 0.9114 & 0.6517 & 0.2327 & 0.1156 \\
(01)(2)(34)                & 0.9075 & 0.6737 & 0.2211 & 0.1053 & 0.9095 & 0.6528 & 0.2490 & 0.0983 \\
(012)(34)                  & 0.9079 & 0.6377 & 0.3623 & 0.0000 & 0.9046 & 0.6585 & 0.3415 & 0.0000 \\ \hline
\end{tabular}
\end{table}

To better globally capture the degree of difference between third-party and user satisfaction annotations, we propose two metrics, agreement ratio (AR) and mean distance (MD). Eq~\ref{eq:agreement ratio} and Eq~\ref{eq:mean distance} show the definition of AR and MD, respectively. 

\begin{equation}
	AR = \dfrac{\sum_{d\in D} I(r(s_u^d) = g(r(S_A^d)))}{|D|}
	\label{eq:agreement ratio}
\end{equation}

\begin{equation}
	MD = \dfrac{1}{D}\sum_{d\in D} |r(s_u^d) - g(r(s_A^d))|
	\label{eq:mean distance}
\end{equation}
In the equations, $r(s)$ denotes the reduction strategy adopted for the satisfaction grade $s$, $S_u^d$ and $s_A^d$ denote the satisfaction annotation results of user and third-party under item (dialog or turn) $d$, $g(S)$ denotes the aggregation strategy adopted for the satisfaction grade set $S$. In our experiment, $r(s) = \dfrac{1}{|R_s|} \sum_{t \in R_s} t$, where $R_s$ is the reduced satisfaction group grade $s$ belongs to, function $g(S)$ is set as the median function. Table~\ref{tab:agreement ratio and mean distance between third-party annotations and user's own annotations} shows the values of AR and MD between third-party annotations and user's annotations using different reduction strategies at different annotation stages. The results demonstrate the facilitative effect of the revision and consultation process. As the annotation stage proceeds, the distance between third-party annotation and user annotation decreases, and the percentage of their agreement rises to some extent. At the session-level, the aggregation of the senior assessor (\textit{step3-2}) is more consistent with the user's annotations than the result of discussion (\textit{step3-1}). This is probably a piece of good news. After all, compared to \textit{step3-1}, \textit{step3-2} results in a much lower experimental overhead. Table~\ref{tab:agreement ratio and mean distance between third-party annotations and user's own annotations} also shows a higher consistency in turn-level annotation compared with the session-level one. This may be caused by the more complicated scenario faced by session-level annotation.

\begin{table}[]
\caption{The agreement ratio (AR) and mean distance (MD) between third-party annotations and user's own annotations on different reduction strategies and different annotation stages.}
\label{tab:agreement ratio and mean distance between third-party annotations and user's own annotations}
\begin{tabular}{c|cccc|cccc}
\hline
\multirow{3}{*}{stage} & \multicolumn{4}{c|}{session-level}                                                & \multicolumn{4}{c}{turn-level}                                                  \\ \cline{2-9} 
                       & \multicolumn{2}{c|}{(0)(1)(2)(3)(4)}        & \multicolumn{2}{c|}{(0)(1)(2)(34)} & \multicolumn{2}{c|}{(0)(1)(2)(3)(4)}        & \multicolumn{2}{c}{(0)(1)(2)(34)} \\ \cline{2-9} 
                       & AR & \multicolumn{1}{c|}{MD} & AR       & MD       & AR & \multicolumn{1}{c|}{MD} & AR       & MD      \\ \hline
orig                   & 0.3333      & \multicolumn{1}{c|}{0.9308}   & 0.6164            & 0.4906         & 0.3871      & \multicolumn{1}{c|}{0.8387}   & 0.6833            & 0.4326        \\
step1                  & 0.3145      & \multicolumn{1}{c|}{0.9245}   & 0.6164            & 0.4843         & 0.4208      & \multicolumn{1}{c|}{0.8079}   & 0.6965            & 0.4106        \\
step2                  & 0.3585      & \multicolumn{1}{c|}{0.8868}   & 0.6415            & 0.4654         & 0.4208      & \multicolumn{1}{c|}{0.7977}   & 0.6950            & 0.4106        \\
step3-1                & 0.3459      & \multicolumn{1}{c|}{0.8868}   & 0.6289            & 0.4717         & -           & \multicolumn{1}{c|}{-}        & -                 & -             \\
step3-2                & 0.3774      & \multicolumn{1}{c|}{0.8491}   & 0.6604            & 0.4465         & 0.4150      & \multicolumn{1}{c|}{0.8021}   & 0.6848            & 0.4194        \\ \hline
\end{tabular}
\end{table}

\subsection{Pairwise analysis}
Considering that the standard of annotation may vary among different users, we further carry out a pairwise analysis. We assume that there is a clear preference between the items with different grades of satisfaction given by the same user. To conduct pairwise consistency evaluation, we propose two methods: MP (first merge then pairwise) and PM (first pairwise then merge). In MP method, we first merge the annotation results of third-party assessors, and then calculate the ratio of  consistency pairs between user and third-party. While in PM method, we first record the preference information of each third-party assessor, and then merge them to calculate the consistent ratio. For example, the third-party annotation results of item $A$ and $B$ is 3-2-3 and 3-3-4, respectively. In MP, the merged annotation results of item $A$ and $B$ are both grade $3$, the model considers them the equivalent level of satisfaction. In PM, the preferences of third-party assessors are $A = B$, $A \prec B$ and $A \prec B$ respectively, then the model considers item $B$ is better than item $A$.

\begin{table}[]
\caption{The agreement ratio of preference pairs between the annotations of third-party and user's own. The table shows results of different aggregation methods and different annotation stages in both session-level and turn-level.}
\label{tab:The agreement ratio of preference pairs between the annotations of third-party and user's own}
\begin{tabular}{c|cccc|cccc}
\hline
\multirow{3}{*}{stage} & \multicolumn{4}{c|}{session-level}                                         & \multicolumn{4}{c}{turn-level}                                           \\ \cline{2-9} 
                       & \multicolumn{2}{c|}{(0)(1)(2)(3)(4)} & \multicolumn{2}{c|}{(0)(1)(2)(34)} & \multicolumn{2}{c|}{(0)(1)(2)(3)(4)} & \multicolumn{2}{c}{(0)(1)(2)(34)} \\ \cline{2-9} 
                       & MP     & \multicolumn{1}{c|}{PM}     & MP               & PM              & MP     & \multicolumn{1}{c|}{PM}     & MP              & PM              \\ \hline
orig                   & 0.7143 & \multicolumn{1}{c|}{0.7669} & 0.7525           & 0.7877          & 0.7416 & \multicolumn{1}{c|}{0.7448} & 0.8078          & 0.7998          \\
step1                  & 0.6907 & \multicolumn{1}{c|}{0.7199} & 0.7549           & 0.7445          & 0.7499 & \multicolumn{1}{c|}{0.7655} & 0.8384          & 0.8452          \\
step2                  & 0.6780 & \multicolumn{1}{c|}{0.7443} & 0.7208           & 0.7441          & 0.7509 & \multicolumn{1}{c|}{0.7697} & 0.8416          & 0.8516          \\
step3-1                & 0.7072 & \multicolumn{1}{c|}{0.7166} & 0.7309           & 0.7238          & -      & \multicolumn{1}{c|}{-}      & -               & -               \\
step3-2                & 0.7185 & \multicolumn{1}{c|}{0.7185} & 0.7502           & 0.7502          & 0.7165 & \multicolumn{1}{c|}{0.7165} & 0.7207          & 0.7207          \\ \hline
\end{tabular}
\end{table}

Table~\ref{tab:The agreement ratio of preference pairs between the annotations of third-party and user's own} shows the agreement ratio of preference pairs between the annotations of third-party and user. The results show that the revision process leads to little improvement, or even a slightly negative impact, in consistency between third parties and users from the pairwise perspective. In particular, conducting \textit{step3-2} at turn-level causes a large decrease in consistency. This phenomenon contradicts the findings of Sec~\ref{subsubsec:Pointwise analysis}. We speculate that it is due to the noise existing in the annotation results. In pairwise mode, the evaluation metric of consistent ratio will amplify the influence of popular-user-generated items with significant differences between the annotations of user and third-party. The results also show that third-party satisfaction annotations at the turn-level are more likely to be consistent with the user's annotations than those at the session-level. This finding is in line with Sec~\ref{subsubsec:Pointwise analysis}.

\subsection{Case analysis}
To better understand the factors that contribute to the difference in the perceptions of satisfaction between users and third-party annotators, we conduct case studies on $33$ dialogs where the difference of satisfaction rating between user and third-party is no less than $2$. The evidence for the analysis is based on the dialog improvement suggestions (optional) given by users in the post-dialog annotation and the text notes and audio records collected during the third-party revision and consultation process.

\begin{table}[]
\caption{The reasons for a low rating on the dialogs where the difference of satisfaction rating between the user and third-party is no less than $2$.}
\label{tab:The reasons for low rating with the dialogs}
    \begin{tabular}{p{2cm}<{\centering}|p{10cm}|p{1cm}<{\centering}}
    \hline
    low   rating by              & reasons for low   rating                                                                                                                                                                  & \#dialogs \\ \hline
    \multirow{10}{*}{user}       & 1.   Agent response is too short to contain sufficient useful information.                                                                                                                & 8         \\
                                 & 2. Agent response is too lengthy to exhaust the effort in extracting useful information.                                                                                                & 3         \\
                                 & 3. Agent response is weakly relevant and does not answer the information needs.                                                                                                         & 2         \\
                                 & 4. Agent response is too unilateral and subjective.                                                                                                                                     & 2         \\
                                 & 5. The Agent is too slow to reply.                                                                                                                                                            & 2         \\
                                 & 6. The tone of the agent's response is too aggressive.                                                                                                                                        & 1         \\
                                 & 7. The information given in the agent's response is out of date.                                                                                                                          & 1         \\
                                 & 8. The organization and presentation of information in agent response are too chaotic.                                                                                                  & 1         \\
                                 & 9. Agent response lack of appropriate clarifying questions to narrow down user information needs                                                                                         & 1         \\
                                 & 10. user does not provide the reason for the low rating                                                                                                                                     & 8         \\ \hline
    \multirow{3}{*}{third-party} & 11.   Agent response is too short to contain sufficient useful information.                                                                                                               & 2         \\
                                 & 12. Agent response is weakly relevant and does   not answer the information needs.                                                                                                        & 1         \\
                                 & 13. In essence, the user also perceives the   agent response as low quality (giving consistent improvement suggestions with third-parties), but he/she still scores it with a high grade. & 1         \\ \hline
    \end{tabular}
    \end{table}

Table~\ref{tab:The reasons for low rating with the dialogs} summarizes the reasons for the low rating situations from the perspective of the lower-rating party. Except for the $8$ dialogs where users do not provide any improvement suggestions, it can be observed that the variation in patience with agent response length is the principal reason for the difference between user and third-party annotation. A too-length or too-short response is likely to undermine the user-perceived satisfaction of the dialog. In addition, third-party assessors fail to capture some signals of user-perceived dissatisfaction in some dialogs, such as unilateral, impolite or out-of-date agent responses.

To continue answering RQ1, we find that third-party satisfaction annotation and users' satisfaction annotation can achieve somewhat consistency, but there still exist considerable differences between the two in quite a few dialogs and turns satisfaction annotation due to the variability in the perception of subjective factors such as lengthy or short responses. This finding is consistent with the results of \cite{fu2022evaluating}. The revision and consultation process of third-party annotations improves the consistency between third-party annotation and the user's annotation in terms of absolute satisfaction ratings, but the effect is rather weak in terms of the preference relationship. Moreover, compared to the annotations at the session-level, third-party satisfaction annotations at the turn-level can achieve a higher degree of consistency with the user's annotations.

\section{Satisfaction modeling from turn-level to session-level}
\label{subsec:from turn-level to session-level}

In Sec~\ref{subsec:consistency between third-party annotation and user annotation}, we find that third-party satisfaction annotations can reach a higher consistency with the user's annotations at the turn-level compared with the consistency at the session-level. An interesting question is whether we can use turn-level third-party satisfaction information to model users' session-level satisfaction. If feasible, in the future we are likely to use an automatic system to model turn-level satisfaction so as to evaluate user session-level satisfaction.

\subsection{Evaluation approach}
Considering the variation in the sensitivity of satisfaction annotation by different users, we design a preference prediction task to evaluate the performance of session-level satisfaction modeling. Each sample contains two distinct satisfaction rating dialogs from the same user. Models need to identify which dialog has a higher satisfaction score by modeling their satisfaction scores. Prediction accuracy is used as the metric to evaluate model performance. In order to reduce the noise in the model evaluation process, the following settings are applied when constructing the test set:

\begin{enumerate}
	\item As the ground truth used for the evaluation, the user's own session-level satisfaction results are reduced to the binary setting (012)(34). We set the threshold between the grades $2$ and $3$ essentially because of the dialog improvement suggestions received from users in the post-annotation. Users tend to give substantial deficiencies under dialogs with a rating less than or equal to $2$, while they often simply present thanks-like messages or minor improvement suggestions under dialogs with a rating of $3$ or $4$.
	\item To control the impact of a single dialog on the global performance, we set the maximum time that each dialog can be sampled to $L$. We set $L=10$ in our experiments as a tradeoff between the sample size and the impact of a single sample.
	\item To reduce the bias caused by test data sampling, we carry out $10$ independent test data sampling tasks using $10$ different random seeds. The follow-up reported model accuracy is the mean value of the performance on these $10$ test data.
\end{enumerate}

Following the aforesaid settings, we generate $10$ independent sets of test data, each containing $767$ samples.

In addition to our self-constructed ConvSearch dataset, we also investigate some popular conversational search datasets, such as MultiWOZ~\cite{budzianowski2018multiwoz}, Topical-Chat~\cite{gopalakrishnan2019topical}, MSDialog~\cite{qu2018analyzing}, DailyDialog~\cite{li2017dailydialog}, PersonaChat~\cite{zhang2018personalizing}, WISE~\cite{ren2021wizard} and etc. We find that only Topical-Chat includes both turn-level and session-level satisfaction annotations from the user. However, Topical-Chat is also not appropriate for our evaluation dataset because its data collection setting significantly differs from the conversational search scenario.

Therefore, we only use the ConvSearch dataset as the evaluation dataset. We report the performance of the session-level user satisfaction prediction model based on turn-level third-party satisfaction (ConvSearch-Third) and turn-level user satisfaction (ConvSearch-User), respectively. 

We use the unweighted-median model (introduced in Sec~\ref{subsubsec:statistics models}) as the baseline. In ConvSearch-Third, we also use the third-party session-level satisfaction annotation results under different annotation stages as comparisons.

\subsection{Factor extraction}
\label{subsubsec:Useful factor extraction}
In this section, we intend to analyze and extract some valuable factors at both the turn-level and session-level. These factors guide the construction of session-level satisfaction prediction models in subsequent sections.

At the turn-level, we attempt to investigate which factors affect the weight of turn-level satisfaction in predicting session-level satisfaction. To probe this issue, we consider the fundamental hypothesis that session-level satisfaction can be obtained by a linear weighting of turn-level satisfaction, just as shown in Eq~\ref{eq:linear model hypothesis}: 

\begin{equation}
	\hat{S}_T = \sum_{t\in T} w(t, \Theta_t) s_t,
\label{eq:linear model hypothesis}
\end{equation}

where $t$ and $T$ denote one turn and the turn set of dialog $T$, $s_t$ and $\hat{S}_T$ denote turn-level satisfaction of turn $t$ and the predicted session-level satisfaction of dialog $T$ respectively.

We use the exponential quadratic function as the weighted function $w(t, \Theta_t)$ to cover all the possible trends of the features $\Theta$, including monotonically increasing, monotonically decreasing, increasing and then decreasing, decreasing and then increasing. Eq~\ref{eq:factor quadratic weighted model} shows the specific regression model with turn-level feature $\theta_t$. Each time, we only consider one single feature, so as to investigate the impact magnitude of different features on the weight of turn-level satisfaction. Unlike directly using $ax^2+bx+c$ to model quadratic function, the parameters $a$, $\tau$ and $\lambda$ in Eq~\ref{eq:factor quadratic weighted model} convey substantive meanings. These three parameters indicate convexity, inflection point, and variance term weight respectively.

%
%
\begin{equation}
	\hat{S}_T = \sum_{t\in T} \left[ \lambda \dfrac{\exp{ a(\theta_t-\tau)^2}}{\sum_{u\in T} a(\theta_u-\tau)^2} + (1-\lambda) \dfrac{1}{|T|} \right] s_t
\label{eq:factor quadratic weighted model}
\end{equation}

\begin{figure}[thbp]
   \captionsetup[subfigure]{justification=centering}
   \centering
	\subfloat[turn position]{
     \includegraphics[width=0.45\textwidth]{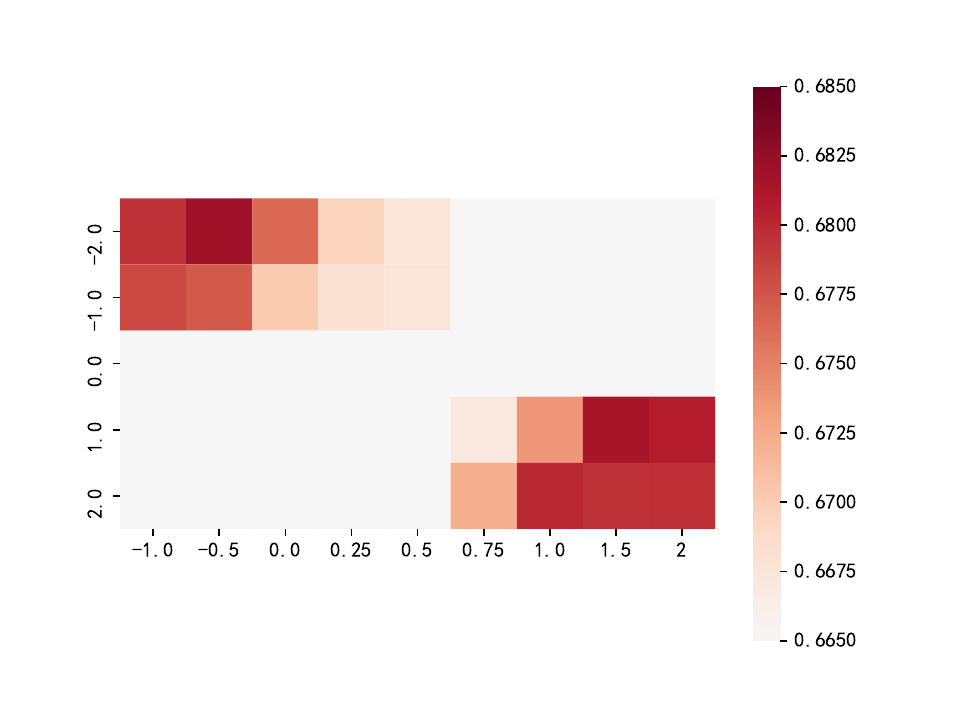}
     \label{fig:turn position, factor}
   }
   	\subfloat[turn satisfaction]{
     \includegraphics[width=0.45\textwidth]{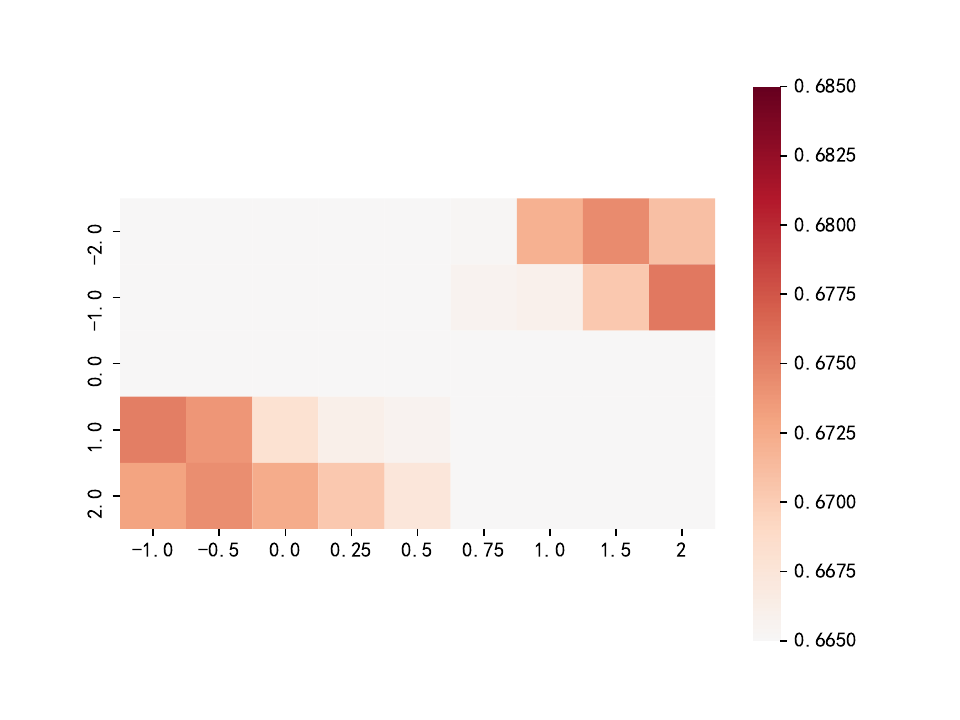}
     \label{fig:turn satisfaction, factor}
   } \\
   \subfloat[agent response length]{
     \includegraphics[width=0.45\textwidth]{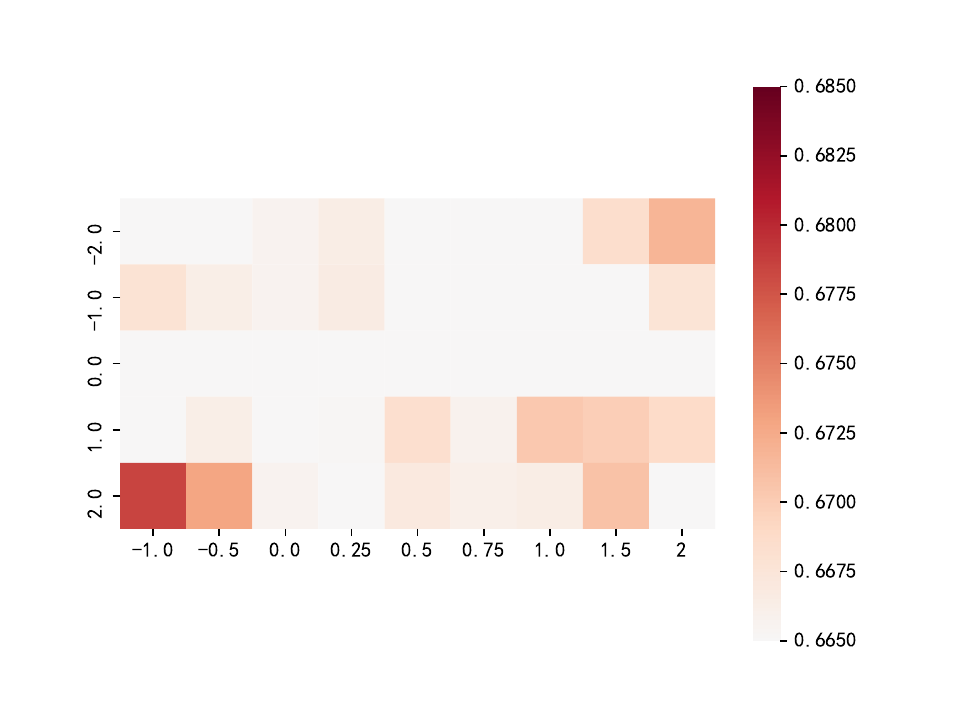}
     \label{fig:agent response length, factor}
   }
   \subfloat[random feature]{
     \includegraphics[width=0.45\textwidth]{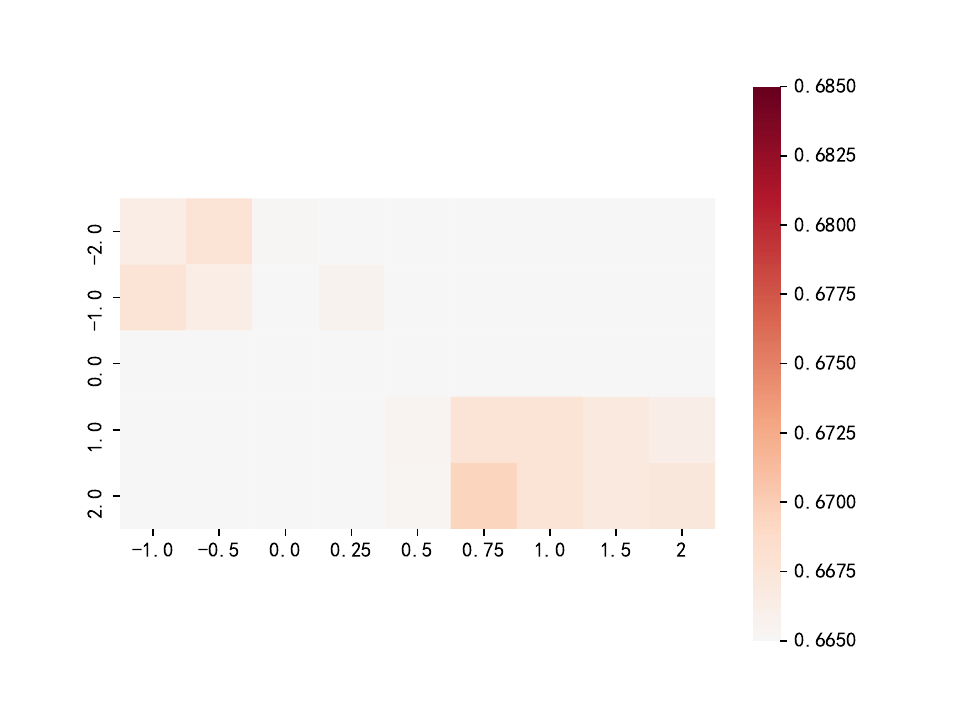}
     \label{fig:random feature, factor}
   }
   \caption{The model performance under quadratic regression (Eq~\ref{eq:factor quadratic weighted model}) with different parameters and different features. In each heatmap, the horizontal and vertical coordinates denote $\tau$ and $a$, respectively.}
   \label{fig:The model performance under quadratic regression with different parameters and different features}
\end{figure}

We separately consider turn position, turn satisfaction and agent response length as features, and then feed them into Eq~\ref{eq:factor quadratic weighted model} for modeling. Fig~\ref{fig:The model performance under quadratic regression with different parameters and different features} shows the model performance with different parameters and different features. In order to eliminate the performance gain due to the variance of parameter selection, we also adopt random features for modeling, as shown in Fig~\ref{fig:random feature, factor}. 

The results show that all selected features achieve higher performance improvements than the random one. Among these three features, the turn position plays the most important role. We find that the model performs better when it assigns higher weights to the dialog-front turns. This phenomenon indicates that users place more focus on the top turns when judging session-level satisfaction. The satisfaction rating of turn-level also affects its importance in predicting session-level satisfaction. Fig~\ref{fig:turn satisfaction, factor} shows that the higher satisfaction ratings weigh more heavily, probably because third-party assessors label the results of low ratings (grade $0$, $1$ or $2$) with less credibility. The magnitude of dissatisfaction might be influenced by personal subjective and emotional factors. As for the agent response length, we find that the model performance can be improved to some extent when the weight of turn is monotonically increasing in some cases or decreasing in others with the increasing response length. This phenomenon suggests that in some dialogs the satisfaction information of long responses is important, while that of short responses is important in other dialogs. We suspect it might come down to the fact that a too-long or too-short response is likely to harm the user's satisfaction of the dialog.

\begin{figure}[thbp]
   \captionsetup[subfigure]{justification=centering}
   \centering
	\subfloat[unsatisfaction ratio in different short turn position]{
     \includegraphics[width=0.45\textwidth]{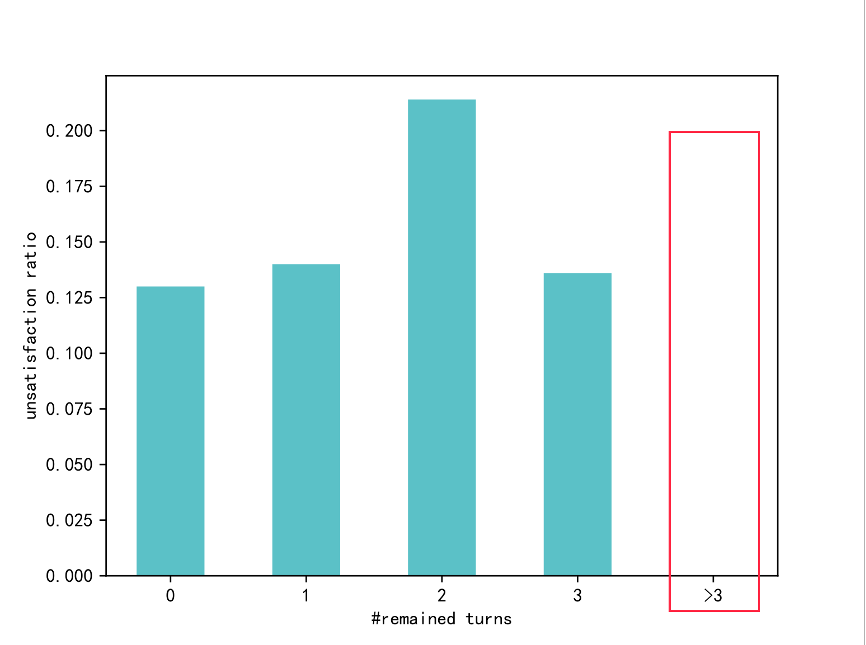}
     \label{fig:unsatratio_remained_turns}
   }
   	\subfloat[dialog length with different session-level satisfaction ratings]{
     \includegraphics[width=0.45\textwidth]{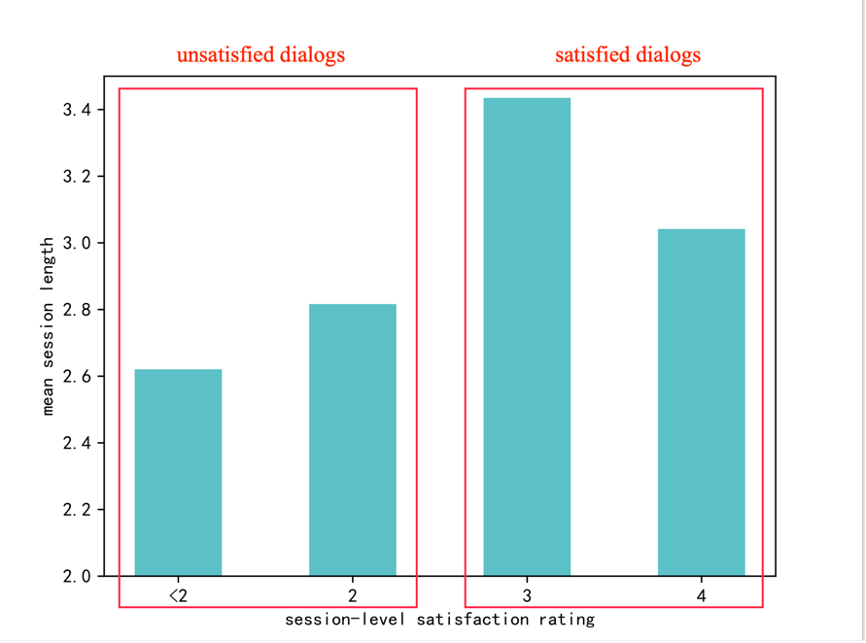}
     \label{fig:dialoglen_satisfaction}
   }
   \caption{Analysis results of session-level features. Fig~\ref{fig:unsatratio_remained_turns} shows the ratio of unsatisfactory dialogs under different numbers of remained turns following a short agent response turn (no more than $20$ words). Fig~\ref{fig:dialoglen_satisfaction} shows  the mean dialog length under different satisfaction ratings.}
   \label{fig:Some session-level findings}
\end{figure}

At session-level, we perform a series of analyses and discover the following two main findings:

\begin{enumerate}
	\item Short turn position finding: Fig~\ref{fig:unsatratio_remained_turns} demonstrates the relationship between the position of short agent responses (no more than $20$ words) and the proportion of session-level dissatisfaction. Result shows that all dialogs are satisfied for those with no less than 4 turns following a short response. This finding illustrates that the short response content in this case does not discourage users from engaging with the conversation.
	\item Dialog length finding: Fig~\ref{fig:dialoglen_satisfaction} shows the  mean number of agent turns per dialog under different user's session-level satisfaction. Huge differences can be found between unsatisfied (grade $0\sim 2$, mean length: $2.62$) and satisfied (grade $3$ or $4$, mean length: $3.15$) dialogs. This is probably because high-quality agent responses could motivate users to stay engaged with the agent. 
\end{enumerate}

\subsection{Statistics models}
\label{subsubsec:statistics models}
We first attempt to model user's session-level satisfaction based on the hypothesis of Eq~\ref{eq:linear model hypothesis}. The key point here is how to model the turn-level satisfaction $s_t$ and its weight $w(t, \Theta)$. 

For $s_t$, in ConvSearch-Third, we need to aggregate the annotation results of different third-party annotators; and in ConvSearch-User, we need to aggregate the turn-level satisfaction information of consecutive turns of agent responses annotated by the user. Specifically, we adopt the following five aggregation methods:

\begin{enumerate}
	\item median/mean/min: use median/mean/minimum of annotation results as the aggregation function. 
	\item interpolate: use x.5 to replace the mean value of annotation results if it is not integer, otherwise use the mean value directly as the aggregation result. 
	\item rescore: first map each of grade $0$, $1$, $2$, $3$ and $4$ to a new satisfaction score, and then use the mean value of these mapped scores as the aggregated results.  The results of grid search show that using the mapping of $1$, $4$, $4$, $20$, $60$ is the optimum, so the performance of the rescore method will be reported following this mapping approach.
\end{enumerate}

For $w(t, \Theta)$, the results of Sec~\ref{subsubsec:Useful factor extraction} show that turn position is the most effective features we have explored, so here we only consider it in the modeling of the $w(t, \Theta)$ function. We still use Eq~\ref{eq:factor quadratic weighted model} to model the $w$ function with turn position as the parameter $theta_u$ (weighted mode). For a baseline comparison, we also consider to set equivalent weights for turn-level satisfaction (unweighted mode), i.e., $w = 1/|T|$.

In addition, we also try to integrate the session-level findings of Sec~\ref{subsubsec:Useful factor extraction} into statistic models as supervised constraints. For the short turn position finding (\textit{+short} tag), we set the maximum score of $s_t$ to be $4$ for the dialogs of which the number of remaining turns following the short response is no less than $4$, or it is set to be $3$ for the other dialogs. For the dialog length finding (\textit{+dlen} tag), we add the dialog length term into the predicted session-level results with an appropriate weighting when the original predicted ratings in the sample pair are both high and close.

\subsection{The DCEBM models}
\label{subsubsec:dialog behavior model}
The previous analysis reveals that the continuing or ending dialog behaviors of the user are great important signals, which may be able to make up for the deficiency of solely using turn-level satisfaction information. In this section, we propose a \textbf{D}ialog \text{C}ontinuation or \textbf{Ending} \textbf{B}ehavior \textbf{M}odel (DCEBM) to integrate these user behavior signals.

In the DCEBM model, we decompose the user's continuation or ending behavior into two implicit factors: patience ($p_t$) and satisfaction ($s_t$). Patience ($p_t$) depicts the probability that the user still has the patience to continue the conversation after the current turn $t$. Satisfaction ($s_t$) depicts the probability that the user has been satisfied with enough information obtained after the current turn $t$. The basic assumption of our model is that the user is willing to continue the conversation if and only if he is still patient and not already satisfied, just as Eq~\ref{eq:dialog behavior model basic assumption} shows.

\begin{equation}
	P(t \neq |T|) = p_t (1 - s_t)
\label{eq:dialog behavior model basic assumption}
\end{equation}

We assume that $p_t$ and $s_t$ are both affected by the information of all previous turns. The influence of the previous turn on $s_t$ decays with time but $p_t$ does not. Eq~\ref{eq:definition of st and pt} shows the specific definitions of $s_t$ and $p_t$:

\begin{align}
	s_t &= \dfrac{s_0 \exp{\left(\sum_{u<t} \alpha(u, t) f(\Theta^s_u) \right)}}{1 + s_0 \exp{\left(\sum_{u<t} \alpha(u, t)  f(\Theta^s_u) \right)}} \nonumber \\
    p_t &= \dfrac{p_0 \exp{\left(\sum_{u<t} g(\Theta^p_u) \right)}}{1 + p_0 \exp{\left(\sum_{u<t} g(\Theta^p_u) \right)}},
\label{eq:definition of st and pt}
\end{align}

where $\Theta^s$ and $\Theta^p$ is the turn-level features used in $s_t$ and $p_t$ respectively. Here we consider three turn-level features: turn-level satisfaction, turn position and content length. $\alpha(u, t)$ denotes the time-decay term. We set $\alpha(u, t) = e^{-\gamma (t-u)}$.

The optimization objective of the DCEBM model is to maximize the likelihood function of fitting the user's continuation or ending conversation behaviors. The likelihood function is shown in Eq~\ref{eq:dialog behavior model likelihood function}, where $D$ and $T_D$ denote one dialog and turn set of this dialog respectively. We directly adopt the stochastic gradient descent (SGD) algorithm~\cite{bottou2012stochastic} to optimize it.

\begin{equation}
	L = \Pi_{D} \left\{ \left(1-p_{T_D}(1-s_{T_D}) \right) \Pi_{t < T_D} p_t (1-s_t) \right\}
\label{eq:dialog behavior model likelihood function}
\end{equation}

During model inference, we consider that the predicted session-level satisfaction is dependent on the satisfaction $s_t$ of all moments and the final patience $p_T$, as Eq~\ref{eq:the definition of predicted session-level satisfaction on dialog behavior model} shows.

\begin{equation}
	\hat{S}_T = h(p_T, s_1, s_2, ..., s_T) = (1-\lambda) p_T + \lambda \dfrac{1-\beta^T}{1-\beta}\sum_{t=1}^{|T|} s_t\ \beta^{t-1}
\label{eq:the definition of predicted session-level satisfaction on dialog behavior model}
\end{equation}

In our modeling, we employ the following two kinds of settings for $\Theta^s$ and $\Theta^p$:

\begin{itemize}
	\item sat+tlen: we set turn-level satisfaction and turn's content length as the features for both $\Theta^s$ and $\Theta^p$.
	\item sat+tlen+posi: we set turn-level satisfaction and turn's content length as the features for $\Theta^s$, while turn's content length and turn position for $\Theta^p$.
\end{itemize}

In order to avoid overfitting during parameter learning, we set some anchor points for each feature and divided the value of feature into several classes for discretized modeling. Specifically, feature of turn-level satisfaction is divided into $[0, 2]$, $(2, 3)$, $3$ and $(3, 4]$ classes. Feature of turn's content length is divided into $[0, 50]$, $(50, 500]$ and $(500, +\infty)$ classes. Feature of turn position is divided into $1$, $2$, $3$, $4$ and $[5, +\infty)$ classes.

\subsection{Model performance}
\label{subsubsec:model performance}

\begin{table}[]
\caption{The prediction performance of different models using third-party turn-level satisfaction, user turn-level satisfaction and third-party session-level satisfaction respectively. * and ** denote the $p$-value of one-side Cochran-Cox test~\cite{cochran1950experimantal} compared with median-unweighted model is larger than $0.9$ and $0.95$, respectively.}
\label{tab:The prediction performance of different models using third-party turn-level satisfaction, user turn-level satisfaction and third-party session-level satisfaction respectively}
    \begin{tabular}{ccccccc}
    \hline
    \multicolumn{3}{c|}{\multirow{2}{*}{model}}                                                                                                          & \multicolumn{2}{c|}{ConvSearch-Third}                                     & \multicolumn{2}{c}{ConvSearch-User}                         \\ \cline{4-7} 
    \multicolumn{3}{c|}{}                                                                                                                                & \multicolumn{1}{c|}{unweighted}        & \multicolumn{1}{c|}{weighted}          & \multicolumn{1}{c|}{unweighted}      & weighted             \\ \hline
    \multicolumn{1}{c|}{\multirow{9}{*}{turn}}   & \multicolumn{1}{c|}{\multirow{7}{*}{statistics models}}      & \multicolumn{1}{c|}{median}             & \multicolumn{1}{c|}{0.6461}            & \multicolumn{1}{c|}{0.6714}            & \multicolumn{1}{c|}{0.8633}          & 0.8739               \\
    \multicolumn{1}{c|}{}                        & \multicolumn{1}{c|}{}                                       & \multicolumn{1}{c|}{mean}               & \multicolumn{1}{c|}{0.6651}            & \multicolumn{1}{c|}{0.6804}            & \multicolumn{1}{c|}{0.8656}          & 0.8775               \\
    \multicolumn{1}{c|}{}                        & \multicolumn{1}{c|}{}                                       & \multicolumn{1}{c|}{interpolate}        & \multicolumn{1}{c|}{0.6522}            & \multicolumn{1}{c|}{0.6846}            & \multicolumn{1}{c|}{0.8678}          & 0.8813               \\
    \multicolumn{1}{c|}{}                        & \multicolumn{1}{c|}{}                                       & \multicolumn{1}{c|}{min}                & \multicolumn{1}{c|}{0.6569}            & \multicolumn{1}{c|}{0.6701}            & \multicolumn{1}{c|}{0.8636}          & 0.8777               \\
    \multicolumn{1}{c|}{}                        & \multicolumn{1}{c|}{}                                       & \multicolumn{1}{c|}{rescore}            & \multicolumn{1}{c|}{0.6767}            & \multicolumn{1}{c|}{0.6861}            & \multicolumn{1}{c|}{0.8713}          & 0.8782               \\
    \multicolumn{1}{c|}{}                        & \multicolumn{1}{c|}{}                                       & \multicolumn{1}{c|}{rescore+short}      & \multicolumn{1}{c|}{0.6773}            & \multicolumn{1}{c|}{0.6921*}           & \multicolumn{1}{c|}{0.8715}          & 0.8787               \\
    \multicolumn{1}{c|}{}                        & \multicolumn{1}{c|}{}                                       & \multicolumn{1}{c|}{rescore+short+dlen} & \multicolumn{1}{c|}{0.6842}            & \multicolumn{1}{c|}{\textbf{0.6963**}} & \multicolumn{1}{c|}{\textbf{0.8788}} & \textbf{0.8829}      \\ \cline{2-7} 
    \multicolumn{1}{c|}{}                        & \multicolumn{1}{c|}{\multirow{2}{*}{DCEBM models}} & \multicolumn{1}{c|}{sat+tlen}           & \multicolumn{1}{c|}{0.7001**}          & \multicolumn{1}{c|}{\textbf{-}}         & \multicolumn{1}{c|}{0.8692}          &      \multicolumn{1}{c}{-}                 \\
    \multicolumn{1}{c|}{}                        & \multicolumn{1}{c|}{}                                       & \multicolumn{1}{c|}{sat+tlen+posi}      & \multicolumn{1}{c|}{\textbf{0.7012**}} & \multicolumn{1}{c|}{-}                  & \multicolumn{1}{c|}{0.8694}          & \multicolumn{1}{c}{-} \\ \hline
    \multicolumn{1}{c|}{\multirow{5}{*}{dialog}} & \multicolumn{2}{c|}{orig}                                                                      & \multicolumn{2}{c|}{0.6457}                                                     & \multicolumn{2}{c}{-} \\
    \multicolumn{1}{c|}{}                        & \multicolumn{2}{c|}{step1}                                                                            & \multicolumn{2}{c|}{0.6680}                                                     & \multicolumn{2}{c}{-} \\
    \multicolumn{1}{c|}{}                        & \multicolumn{2}{c|}{step2}                                                                            & \multicolumn{2}{c|}{0.6686}                                                     & \multicolumn{2}{c}{-} \\
    \multicolumn{1}{c|}{}                        & \multicolumn{2}{c|}{step3-1}                                                                          & \multicolumn{2}{c|}{\textbf{0.6896*}}                                           & \multicolumn{2}{c}{-}              \\
    \multicolumn{1}{c|}{}                        & \multicolumn{2}{c|}{step3-2}                                                                          & \multicolumn{2}{c|}{0.6713}                                                     & \multicolumn{2}{c}{-} \\ \hline
    \multicolumn{1}{l}{}                         & \multicolumn{1}{l}{}                                        & \multicolumn{1}{l}{}                    & \multicolumn{1}{l}{}                   & \multicolumn{1}{l}{}                   & \multicolumn{1}{l}{}                 & \multicolumn{1}{l}{} \\
    \multicolumn{1}{l}{}                         & \multicolumn{1}{l}{}                                        & \multicolumn{1}{l}{}                    & \multicolumn{1}{l}{}                   & \multicolumn{1}{l}{}                   & \multicolumn{1}{l}{}                 & \multicolumn{1}{l}{}
    \end{tabular}
    \end{table}
    
Table~\ref{tab:The prediction performance of different models using third-party turn-level satisfaction, user turn-level satisfaction and third-party session-level satisfaction respectively} shows the prediction performance of statistic models and the DCEBM models. In the ConvSearch-Third dataset, Table~\ref{tab:The prediction performance of different models using third-party turn-level satisfaction, user turn-level satisfaction and third-party session-level satisfaction respectively} also lists the prediction results by directly using third-party session-level annotations as a comparison.
 
We find that statistics models incorporating session-level findings achieve some performance improvement compared to the original version. This phenomenon suggests that session-level signals are valuable when constructing session-level satisfaction. However, these session-level supervised signals are likely to be dataset-specific and have poor transferability, which is one of the limitations of the statistical model. When comparing the weighted and unweighted models, we find that the weighted statistic models work much better when turn-position is introduced as the weighted factor, which is in line with the findings of Sec~\ref{subsubsec:Useful factor extraction}.

On one hand, We are pleased to find that the DCEBM models significantly outperform the baseline unweighted median statistic model using the ConvSearch-Third dataset, and the sat-tlen+posi one outperforms all the other models. The result shows that the user's continuation or ending behaviors contain a wealth of information. We can improve the effectiveness of session-level satisfaction prediction by appropriately mining and utilizing these behaviors. On the other hand, we find that the DCEBM models do not result in significant performance improvements compared with the baseline unweighted median statistic model using the ConvSearch-User dataset. We assume that this is due to the way we collect user satisfaction assessments. In our setup, the user is required to first annotate all turn-level ratings, and then conduct session-level annotation. This procedure implicitly induces the user to refer to their turn-level satisfaction results when annotating session-level satisfaction. Hence, statistical models are more advantageous in this setting.

Table~\ref{tab:The prediction performance of different models using third-party turn-level satisfaction, user turn-level satisfaction and third-party session-level satisfaction respectively} also reconfirms the effectiveness of the third-party revision and consultation process. Although third-party session-level satisfaction annotation can achieve competitive results, the time-consuming complex revision and consultation process with high cost renders it a more restricted approach compared to turn-level satisfaction annotation.

To answer RQ2, we justify the feasibility of constructing session-level satisfaction from turn-level satisfaction information. Besides turn-level satisfaction, the position and content length of turns are crucial factors to determine the importance of each turn. We also extract dialog length and short turn position in a dialog as valuable session-level signals and show their effectiveness in model construction. In addition, we construct the DCEBM models based on the user's behavior information of continuing or ending the dialog. Results show that the dialog behavior model outperforms all statistical models based on third-party turn-level annotation information.
\section{Discussions}
\label{sec:discussions}

\subsection{Feasibility of the Cranfield paradigm}
\label{subsec:Feasibility of the Cranfield Paradigm}
Our experimental results show that third-party annotations can achieve somewhat consistency with users' satisfaction annotations, but there still exists a considerable gap between them. We assume that this is due to the impracticability of third-party annotators to fully understand the user's context and emotional characteristics. Compared with the user's own satisfaction annotations, third-party satisfaction annotations are more objective, which allows third-party annotators to reach a consensus after several rounds of revisions. In the ConvS scenario, user satisfaction is usually affected by various factors, such as emotion and agent response delay, which makes it unrealistic to perfectly model user satisfaction using the Cranfield paradigm.

The difference between third-party and user satisfaction annotations does not entail the infeasibility of the Cranfield paradigm. Based on the findings in our experiments, here we list some suggestions for conducting third-party user satisfaction annotation: (1) There should not be too many annotation grades, binary or three-grade setting is more appropriate. Excessive annotation grade makes it difficult for third-party annotators to reach a consensus. (2) Turn-level satisfaction annotation conducted by a third-party is more reliable than the session-level one. (3) The annotation reasons from third-party annotators are of great importance as a reference for aggregating their annotation results.

\subsection{Valuable information from user behaviors}
\label{subsec:Valuable information from user behaviors}
The improved performance of the dialog behavior models implies that user behaviors contain quite valuable information in the ConvS scenario. Appropriate use of user behavior information in ConvS can effectively compensate for the shortcomings of contextual semantic modeling, thus better-modeling user satisfaction. Besides the continuation or ending behaviors adopted in our models, many other valuable user behaviors are reserved for further explorations, such as the user's emotional state, browsing historical dialog content behavior and time spent typing questions.
\section{Conclusions}
\label{sec:conclusions}
In this paper, we investigate the feasibility and effectiveness of Cranfield paradigm and user behaviors in ConvS user satisfaction modeling. To support associated research, we build a novel ConvS experimental platform and construct a novel Chinese open-domain conversational search dataset containing rich annotations and search behaviors. We also recruit third-party assessors to conduct a series stages of satisfaction judgments at both turn-level and session-level. Experimental results show a moderate consistency among the original third-party annotations, with a significant increase as revision and consultation proceed. Although the satisfaction annotation of third-party assessors can achieve somewhat consistency with the annotation of user, a considerable gap still exists between these two kinds of annotations in several dialogs and turns due to the some personal preference factors. Based on user's continuation or ending behaviors, we also propose dialog behavior models to model session-level user satisfaction using turn-level information. Experimental results show that the dialog behavior models significantly outperform the baseline statistics model with third-party turn-level satisfaction annotations, which indicates a large exploration potential of user behavior in the evaluation of ConvS.

\bibliographystyle{ACM-Reference-Format}
\bibliography{main.bib}

\end{document}